\documentclass[fleqn,usenatbib]{mnras}

\usepackage{newtxtext,newtxmath}

\usepackage[T1]{fontenc}

\DeclareRobustCommand{\VAN}[3]{#2}
\let\VANthebibliography\thebibliography
\def\thebibliography{\DeclareRobustCommand{\VAN}[3]{##3}\VANthebibliography}

\newcommand{\be}{\begin{equation}}
\newcommand{\en}{\end{equation}}

\def\mgii{Mg~{\sc ii}}

\def\ni2{Ni~{\sc ii}}
\def\nv{N~{\sc v}}

\def\aliii{Al~{\sc iii}}
\def\oiii{O~{\sc iii}}
\def\civ{C~{\sc iv}}
\def\siiv{Si~{\sc iv}}

\def\si2{Si~{\sc ii}}
\def\feii{Fe~{\sc ii}}
\def\h1{H~{\sc i}}
\def\feiii{Fe~{\sc iii}}

\def\kms{km s$^{-1}$}

\usepackage{graphicx}	
\usepackage[dvipsnames]{xcolor}
\usepackage{amsmath}
\usepackage{amssymb} 
\usepackage{adjustbox}
\usepackage{placeins}
\usepackage{float}
\newcommand*{\rom}[1]{\expandafter\@slowromancap\romannumeral #1@}

\title[BAL quasar orientation]{Fraction of broad absorption line quasars in different radio morphologies}
\author[Nair et al.]{
A. Nair,$^{1}$\thanks{E-mail: nairakhilwork@gmail.com}
M. Vivek,$^{2}$\thanks{E-mail: vivek.m@iiap.res.in}
\\
$^{1}$University of Mumbai, Kalina, Mumbai 400 098, India\\
$^{2}$Indian Institute of Astrophysics, Koramangala, Bengaluru 560 034, India\\
}

\date{Accepted XXX. Received YYY; in original form ZZZ}

\pubyear{2021}

\begin{document}
\label{firstpage}
\pagerange{\pageref{firstpage}--\pageref{lastpage}}
\maketitle

\begin{abstract}
In this study, we investigated the orientation model of Broad Absorption Line (BAL) quasars using a sample of  sources that are common in Sloan Digital Sky Survey (SDSS) Data Release (DR)-16 quasar catalog and Very Large Array (VLA)-Faint Images of the Radio Sky at Twenty Centimeters (FIRST) survey.  Using the radio cut-out images from the FIRST survey, we first designed a deep learning model using convolutional neural networks (CNN) to classify the quasar radio morphologies into the core-only, young jet, single lobe, or triples. These radio morphologies are further sub-classified into core-dominated and lobe-dominated sources.    The CNN models can classify the sources with a high precision of >98$\%$ for all the morphological sub-classes. The average BAL fraction in the resolved core, core-dominated, and lobe-dominated quasars are consistent with the BAL fraction inferred from radio and infra-red surveys.  We also present the distribution of BAL quasars as a function of quasar orientation by using the radio core-dominance as an orientation indicator. A similar analysis is performed for HiBALs, LoBALs, and FeLoBALs. All the radio morphological sub-classes and BAL sub-classes show an increase in BAL fraction at high orientation angles of the jets with respect to the line of sight. Our analysis suggests that  BAL quasars are more likely to be found in viewing angles close to the equatorial plane of the quasar. However, a pure orientation model is inadequate, and a combination of orientation and evolution is probably the best way to explain the complete BAL phenomena. 
\end{abstract}




\section{Introduction}
\label{sec:intro}
Quasars are highly luminous radio sources powered by accretion onto supermassive black holes located at the center of their host galaxies. These sources consist of non-thermal and polarized radio emissions, which can be seen as jets arising from the central region of the quasar. Such quasars can have different morphologies in the radio wavelengths depending on the orientation of the radio jets with respect to the line of sight. In quasars with extended radio emissions, sometimes only a single lobe is visible, whereas double lobes along with a core at the center are seen in many other cases. These radio lobes can extend from a few kiloparsecs to megaparsec scales on each side of the central core. Fanaroff \& Riley (FR) classification, based on the brightness of the extended radio sources, consists of low-luminosity FR I and high luminosity FR II sources \citep{Fanaroff1974}.
In FR I sources, radio luminosity becomes fainter as one moves away from the core. Conversely, the end of the radio lobes, also known as 'hot spots', are brighter in FR II sources. As of now, multiple sub-classifications have been defined to describe the morphological structure of these quasars, a few of which include core/lobe-dominated quasars  \citep{Kimball2011}; X-shaped radio galaxies \citep{Cheung2007, Leahy1992}; narrow/wide-angle tail sources \citep{Rudnick1976}; ring-like radio galaxies \citep{Proctor2011}.
 
Outflows from the central regions of active galactic nuclei (AGN) are thought to be the main agents that regulate the evolution of both the central supermassive black holes as well as the host galaxies \citep{Silk1998, Dimatteo2005}.
The presence of high-velocity outflows from AGN can be established from the evidence provided by the blue-shifted broad absorption lines (BALs) seen in the spectra of 10--20\% of quasi-stellar objects (QSOs)\footnote{In this work, we use 'QSO' to refer optically selected AGN and 'quasar' to denote AGN with radio emission.} \citep{Weymann1991}. BALQSOs are classified into three sub-classes based on the ionization state of the absorbing gas:(a) high-ionization BALQSOs (HiBALs) consists of absorption from high-ionization lines such as \civ, \siiv, and \nv\ \citep{Gibson2008,FilizAk2013,Vivek2016,Vivek2019,Mishra2021} (b) low-ionization BALQSOs (LoBALs) show absorption from low-ionization lines such as \mgii\ and \aliii\ along with the high-ionization lines \citep{Voit1993,Vivek2014,Vivek2018,Yi2019}, and (c) iron-LoBALs (FeLoBALs)  are LoBALs with excited fine-structure \feii\ and/or \feiii\ absorption lines \citep{Vivek2012,McGraw2015}. The observed BALQSO fraction is explained either by an orientation model, where the line of sight intersects with the BAL absorbing clouds in 10--20\% QSOs \citep{Weymann1991, Elvis2000}, or an evolutionary model, where the QSO spends 10--20\% of its lifetime in the BALQSO phase \citep{Farrah2007, Lipari2009}. Although the similarity in the optical/NIR properties of BAL and non-BALQSOs,  higher reddening and polarization features in BALQSOs support the orientation model, the observed radio spectral indices of BALQSOs and the discovery of polar winds in BALQSOs challenge this scenario \citep{Reichard2003, Wang2007, Montenegro-Montes2008, Zhou2006}. On the other hand, the anti-correlation between the radio loudness parameter and balnicity index, Compact Steep Spectrum (CSS) nature of BALQSOs, and the redshift dependence of BALQSO fraction favour the evolutionary model \citep{Gregg2006, Farrah2007,Allen2011}.

The exact nature of BALQSOs is yet unsettled mainly due to the lack of a reliable orientation indicator in quasars. The current approach uses statistical indicators of orientation for large data samples that involve properties from the radio, UV, and optical domains. The indicators in the UV/optical region are based on  the properties of emission lines. 
A strong correlation is found between the velocity offsets and line width ratio of the high-ionization \civ\  and the low-ionization \mgii\  lines in the ultraviolet region. Simulations carried out by \cite{Yong2019} show that an increase in the line width ratio of \civ\  and \mgii\  line is observed for face-on viewing angles as compared to the edge-on viewing angles. Properties of emission lines such as the full width at half maximum (FWHM) of H$\beta$ and the equivalent width of \oiii\  have also been studied as an orientation indicator for low redshift quasars \citep{Shen2014,Risaliti2011}.

In the standard AGN unification theory, the anisotropic radio emission produced by relativistic jets points to a description where  properties like radio morphology and radio spectral index depend on orientation.  Two widely used orientation indicators in the radio regime are the (a) core-to-lobe flux density ratio (R), which is the ratio of the flux densities of the core and lobe in radio wavelengths, and (b)  radio-to-optical ratio of the quasar core (R$_I$), which is the ratio of flux densities of the core in radio and optical wavelengths (See Eqn.~\ref{eq:Rpar}\&\ref{eq:R1par} for the exact definition of R and R$_I$ parameters).  
A high R-value suggests a small viewing angle to the jet axis, while a low R-value indicates more of an edge-on view to the quasar \citep{Orr1982, Kapahi1982, Morisawa1987, Morganti1997}. Studies have shown that R correlates with core radio luminosity \citep{Hardcastle2000} as well with the core optical luminosity \citep{Kharb2004}. 
Factors like age and environment are believed to introduce the scatter in the correlation involving the R parameter suggesting that R might not be the best measure for orientation. The R parameter depends on both the core's radio flux as well as the radio flux of the lobes arising from the quasar jets. These jets are suggested to be affected by environmental factors.  \cite{Wills1995}, and \cite{Kharb2010} showed that the radio-to-optical ratio (R$_I$) of the quasar core might be a better measure of studying the orientation. The R$_I$ parameter measures the core-boosting factor by normalizing the core's radio flux by the core's optical flux. Although with some scatter in the correlation, \cite{Kimball2011} demonstrated a strong correlation between R and R$_I$ parameter suggesting that the two parameters are good indicators of quasar orientation. The scatter in the correlation supports the idea that other factors like age, size, environment, and luminosity also influence these two measurements. 

Quasar orientation in BALQSOs has been studied with the help of radio spectral indices of the sources \citep{Montenegro-Montes2008, Fine2011, Bruni2012}.  \citet{Montenegro-Montes2008} found that for a sample of 15 BALQSOs, all sources had convex spectra typical for CSS/GPS sources indicating young age for these sources. However, \citet{Bruni2012} found no difference in the incidence of GPS sources amongst BALQSOs and non-BALQSOs, inconsistent with BALQSOs as a complete class of younger objects. \citet{DiPompeo2011a} showed that BAL quasars have an over-abundance of steep spectrum sources compared to non-BAL quasars, favoring the orientation model. Thus, radio spectral index studies are not yet successful in completely settling the debate between orientation and evolutionary models, most likely due to the small number of sources in these studies. Studying the orientation of the quasar using the R and R$_I$ parameter in large sets of quasars with different radio morphologies is helpful to obtain a better picture of the orientation dependence of BAL quasars. With the availability of large-sky-area high-sensitive radio continuum surveys, such as Very Large Array (VLA)-Faint Images of the Radio Sky at Twenty Centimeters (FIRST) survey \citep{Becker1995}, it is now possible to carry out statistical studies of different morphological classes of radio quasars. Upcoming radio sky surveys such as the Low-Frequency Array (LOFAR) survey \citep{deGasperin2021}, Very Large Array Sky Survey (VLASS) \citep{LacyM2020}, and  Square Kilometer Array (SKA) will increase the sample of radio sources by several multitudes. As a visual classification of these many sources is impractical, machine learning algorithms offer an excellent solution to these big data problems.  Machine learning techniques have already been implemented in many problems such as classification of optical transients \citep{Mahabal2011}, estimation of photometric redshifts \citep{Mountrichas2017}, prediction of solar flares \citep{Benvenuto2018} and even morphological classification of active galactic nucleus \citep{Aniyan2017, Lukic2018}.

In this paper, we built Convolutional Neural Network (CNN) models to classify the SDSS selected radio quasars from the FIRST survey into different radio morphological classes like compact core, young jets, single lobe, and double-lobe. The training sample in this study is derived from the \cite{Kimball2011} catalog, which consists of spectroscopically confirmed quasars with robust visual classifications of radio quasar morphology using FIRST images.  
We use the sources with predicted labels from the CNN model as the base sample for studying the BAL fraction's dependence on the quasar's orientation. Sample of BAL quasars classified as HiBALs, LoBALs, and FeLoBALs from \cite{Trump2006} is also used for checking the orientation dependence in these sub-classes.

This paper is outlined as follows: The quasar sample data  is explained in section \ref{sec:thedata}. In section \ref{sec:cnn} we describe the CNN model used for automatic classification of radio morphologies adopted in our study.  Section \ref{sec:balquasars} discusses  the dependence of BAL quasars on orientation using multiple orientation indicators. Discussion of the results and conclusion are given in sections \ref{sec:discussion} \& \ref{sec:conclusion}.  Throughout the paper, we adopt a cosmology with H\textsubscript{0} = 70 kms$^{-1}$ Mpc$^{-1}$, $\Omega$\textsubscript{M} = 0.3, and $\Omega$\textsubscript{$\Lambda$} = 0.7.
\begin{figure*}
	\includegraphics[scale=0.375]{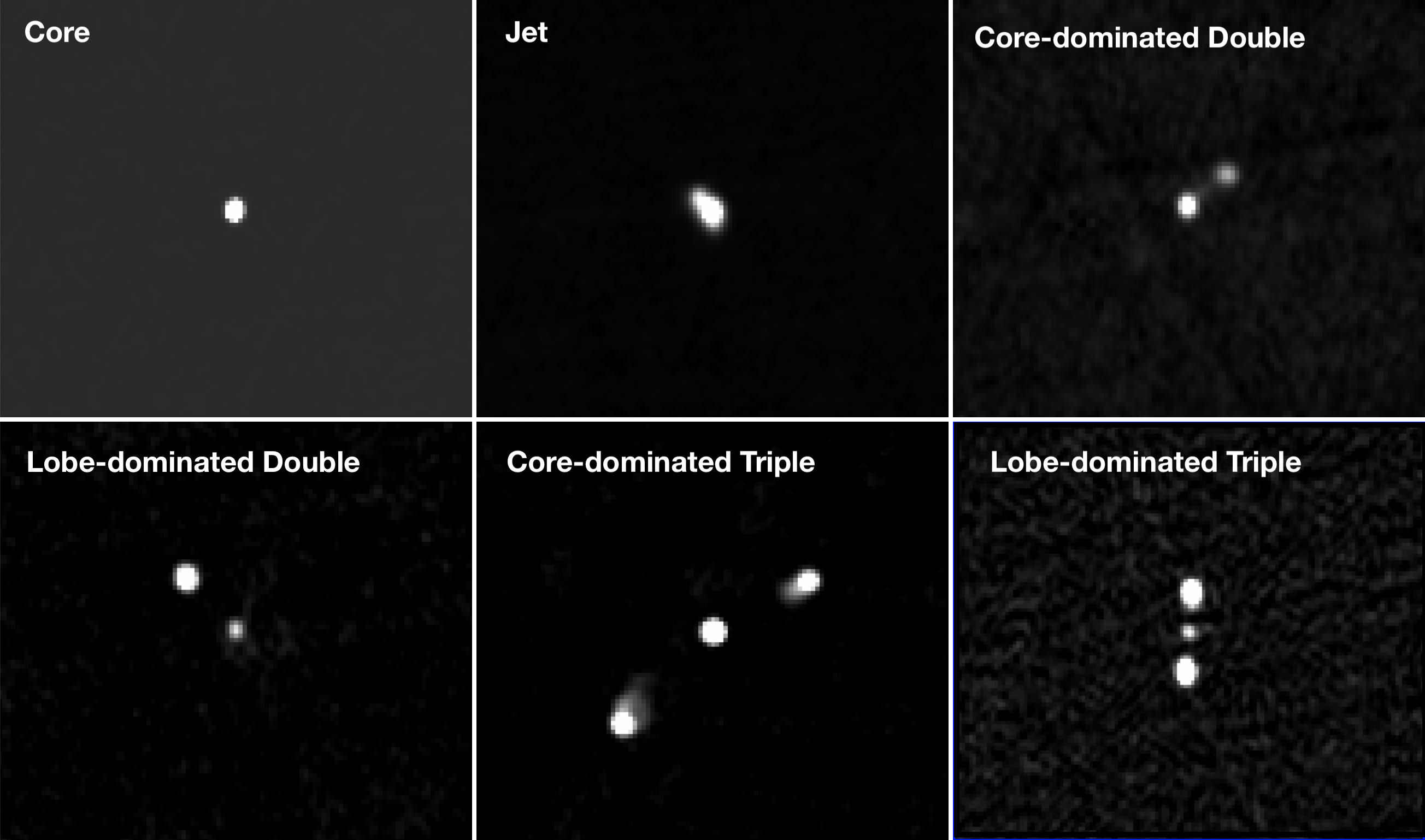}
    \caption{Example cut-out images of sources with different radio morphological classes used in this study. The 4'x4' image cutouts from the FIRST survey are scaled linearly. }
    \label{fig:example_class}
\end{figure*}

\section{The DATA}
\label{sec:thedata}

Our parent quasar sample is drawn from the SDSS DR16 quasar catalog \citep[hereafter DR16Q,][]{Lyke2020}  and the VLA FIRST survey \citep{Becker1995}.  SDSS DR16Q catalog contains the largest set of spectroscopically confirmed quasars. The VLA FIRST survey is the largest available radio continuum survey covering the same region of the sky as the SDSS.  

We first started with the  DR16Q quasar  catalog, containing $\sim$750,414 quasars. This includes 225,082 quasars obtained for the extended Baryon Acoustic Oscillation Spectroscopic Survey \citep[eBOSS,][]{Dawson2016} that appear in the quasar catalog for the first time. The quasar catalog contains around 480,000 sources in the redshift range 0.8 $<$ z $<$ 2.2 and more than 239,000  higher redshift sources observed for the Ly$\alpha$ forest studies. 
Using this catalog, broad absorption line quasars and Damped Ly$\alpha$ systems (DLAs) were classified using automated algorithms by  \citet{Guo2018} and \citet{Parks2018} respectively. The catalog includes  99,856 BALs and 35,686 DLAs. This quasar-only catalog is estimated to be 99.8$\%$ complete with 0.3-1.3$\%$ contamination. BAL quasars are identified using the 'BAL\_PROB' keyword in the DR16Q catalog, which is adapted from \citet{Guo2018} and gives the probability of the quasar to be a BAL quasar. 

The VLA FIRST survey covered $\sim$9055 deg$^{2}$ of the North Galactic Cap and Equatorial Strip. It used the NRAO VLA telescope operating at a frequency of 1.4 GHz (20cm) with a beam size of 5.4," and an RMS sensitivity of about 15 mJy beam$^{-1}$. The catalog contains 946,432 sources with a source detection threshold of $\sim$1 mJy and an astrometric accuracy better than 1". The catalog includes the integrated flux density at 20 cm (S\textsubscript{20}) and peak flux density at 20 cm (S\textsubscript{peak}) for each source. For this study, we make use of the "14dec17" version of the catalog.

{ We first obtained a set of 21,843 sources from the FIRST catalog, which fall within 2" of the SDSS DR16Q catalog. Sources with the integrated flux density at 20 cm \citep[S\textsubscript{20}][]{Kimball2011} > 2mJy were selected to obtain a sample with enough flux to distinguish between different morphological classes. The threshold limit is set not to define a radio-loud population but rather to obtain a sample similar to the training set used for our CNN model. This threshold limit helps the CNN classifier identify and learn the morphological features accurately. We exclude the sources with S\textsubscript{20} $\leq$ 2 mJy from our classification study. The remaining 16,491 "radio sources" are further classified into different radio morphologies using CNN algorithms.}

\subsection{Radio Morphology Classes}
The radio morphology classification used in this study is derived from \cite{Kimball2011} that classified radio sources using a visual examination.  
The sample provided in \cite{Kimball2011} consists of 4714 radio quasars from the VLA FIRST survey. These sources have $S_{20}$ $>$ 2 mJy and an optical counterpart in the SDSS DR7 quasar catalog within 2" of the radio source. These sources are further classified according to their radio morphology into core, doubles, triples, and jets. 
Following \cite{Kimball2011}, we also classify the radio quasars in our sample into the following morphological classes:
\begin{enumerate}
    \item Core; a quasar core with compact radio emission
    \item Jet; a source with a quasar core and young radio jet
    \item Double; a source with radio emission from the quasar core and a single lobe. 
    \item Triple; a double-lobed source with radio emission from quasar core and both lobes
\end{enumerate}

Doubles and Triples are further sub-classified as core-dominated and lobe-dominated radio sources. The core-dominated sources are defined to have a core brighter than its lobe, whereas the lobe-dominated sources have a lobe brighter than its core. \cite{Kimball2011} includes another sub-classification for the triples called "irregular triples", for triples having one lobe brighter than the other one. 
We classify the "irregular triples" as lobe-dominated triples in our classification model. For the classification model, we use image cut-outs of dimension 4'x4' from the VLA FIRST survey, which has an instrumental resolution of ~5". Example cut-out images of sources with different radio morphological classes used in this study are shown in Fig.~\ref{fig:example_class}.

\section{Automatic classification of Radio morphologies}
\label{sec:cnn}

Unlike \citet{Kimball2011}, we use CNN models to automatically classify the radio quasars into different morphological classes.

\subsection{Convolutional Neural Network}

{
 Artificial neural network is a  computational framework inspired by how biological neural networks in the human brain process information. The building block of an artificial neural network is called a neuron, which is responsible for accepting input data, performing calculations, and producing output.  Several neurons are stacked together to form a layer of neurons, and a deep neural network contains many layers placed between the input and output layers. A single neuron receives multiple inputs from all the neurons in the previous layer.  Each of these inputs is multiplied by a weight term, and their sum is sent to neurons in the next layer after applying a bias and an activation function.   The activation function outcome then decides if a neuron is activated or not. An activated neuron transfers information into the other layers. The following equation gives the output of a single neuron,
\begin{equation}
    y= f\left( \sum_{i= 0}^{n} \left(W_i * X_i\right) + B \right)
\end{equation}
where W$_i$, X$_i$, B, n, and f denote the weight vector,  input vector,  bias term, number of neurons, and activation function, respectively. The process of calibrating the values of weights and biases of the network is called training of neural network. 
During the training process, known samples are passed through the network and the network fine-tunes the weights and bias of each neuron based on the error between the predicted responses and known responses in a backpropagation step. Proper tuning of the weights and biases reduces error and makes the model reliable.}

A CNN is built using a series of layers which help in the classification problem. The first layer of an image classifier CNN model  is the input layer, that can be grey-scaled  or  3 channel RGB images. The second type of layer is the convolutional layer, which contain the kernels that  extract features from different classes of images. Kernel convolution is a process in which  a small filter matrix when passed  over the target image transforms the image based on the values of the filter matrix. These filters are tensors which keep track of information and extract features in a convolutional layer. The convolution of the image matrix multiplied with the filter matrix gives the feature map. 
Small kernel sizes help in extracting a large amount of information in cases containing many local features from the input.
Each neuron is then multiplied with an activation function such as the Rectified Linear Unit (ReLU) that introduces the necessary non-linearity into the model. Such non-linearity is essential to produce non-linear decision boundaries. 

Each convolutional layer is followed by a pooling layer which helps in decreasing the dimensionality of the neural network. Max-pooling downsamples the feature map by retaining only the maximum value for patches of a feature map. 
Hence, it contains the most prominent features from the previous feature map. The dropout layer helps to reduce over-fitting by randomly setting the output of neurons to zero for a certain duration of the time during training.  A flatten layer is used to convert the network into a one-dimensional vector. The flattened vector undergoes a few more fully connected dense layers before the output layer.  The  Softmax activation function used in the output layer helps scale the CNN's output into probabilities. It scales the output between 0 and 1. This becomes helpful in the output layer where categorical classification is to be performed.  We implemented the CNN model using the Tensor Flow\footnote{https://www.tensorflow.org}, which is an open-source software for deep learning applications.

\subsection{Pre-Processing of training set}
To train the neural network, we make use of the classified sources from the \cite{Kimball2011} catalog. We collect 4'x4' linearly scaled FIRST image cut-outs of 4517 sources for our training set.

In order to maintain homogeneity in the training set, we use multiple image-transformations to generate additional images from the existing set. Since all the FIRST images obtained from \cite{Kimball2011} have the quasar at the center of the image, the transformations made to generate new images were such that they did not change the position of the central core of the quasar. 
The transformations and their ranges applied to the images are as follows:

\begin{itemize}
    \item Rotation range = 5\textdegree\ and 10\textdegree
    \item Shear range = 0.2
    \item Horizontal flip = True
    \item Vertical flip = True
\end{itemize}

 The rotation transform rotates the image by the specified angle. We also apply a shear transform to the image by providing a shear range. The images are also flipped horizontally and vertically to generate new sets of images.
\begin{table}
	\centering
	\caption{The number of sources in each morphological sub-class, before and after pre-processing of the data.  The first column gives the main class of the source, and the second column gives the sub-classification within that morphological class. The third and fourth columns give the number of sources present before and after the pre-processing, respectively.}
	\label{tab:imgaug}
	\begin{tabular}{l | c | c | c} 
		\hline
		Class & Sub-class & Before  & After \\
		 &  & pre-processing  & pre-processing \\

		\hline
		Core & - & 3433 & 9786\\
		Double$^{\dagger}$ & - & 387 & 9809\\
		Triple$^{\dagger}$ & - & 619 & 9982\\
		Jet & - & 183 & 8437\\
		Double & Core-dominated & 232 & 9038\\
		Double & Lobe-dominated & 150 & 7837\\
		Triple & Core-dominated & 165 & 8171\\
		Triple & Lobe-dominated & 354 & 8360\\
		
		\hline
	\end{tabular}
	\begin{flushleft}
	$^{\dagger}$ 5 doubles and 100 triples are sub-classified as 'unclassified' by \citet{Kimball2011}
	\end{flushleft}
\end{table}
The number of sources before and after applying image-augmentation are shown in Table \ref{tab:imgaug}. A sample with a large number of training data for a specific class will cause the model to over-fit for specific cases. Image-augmentation allows the CNN model to obtain homogeneous input data for all source sub-classes.  We observe a considerable increase in the number of sources with extended radio emissions, which otherwise had few sources.   Such an increase in the data size helps reduce the amount of over-fitting that occurs due to a high number of sources in one specific class.  
\subsection{CNN structure to classify radio quasar morphologies}

\begin{figure*}
\begin{tabular}{c}
   
	\includegraphics[scale=0.40]{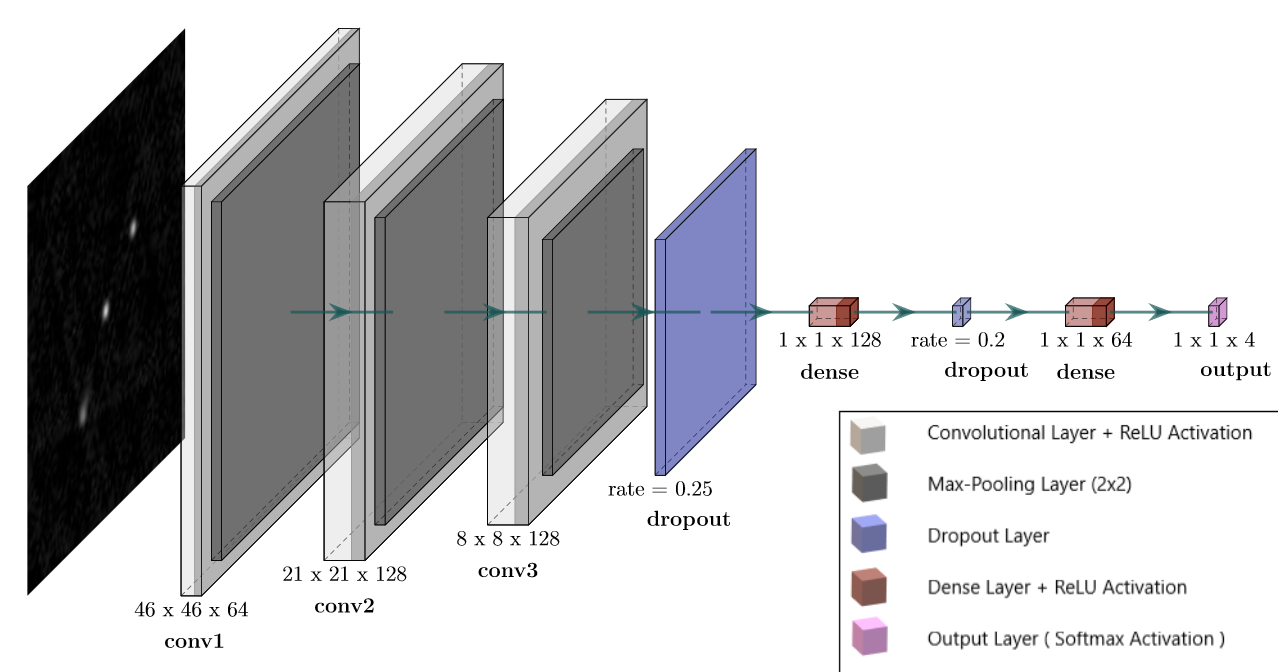}\\
	\includegraphics[scale=0.4]{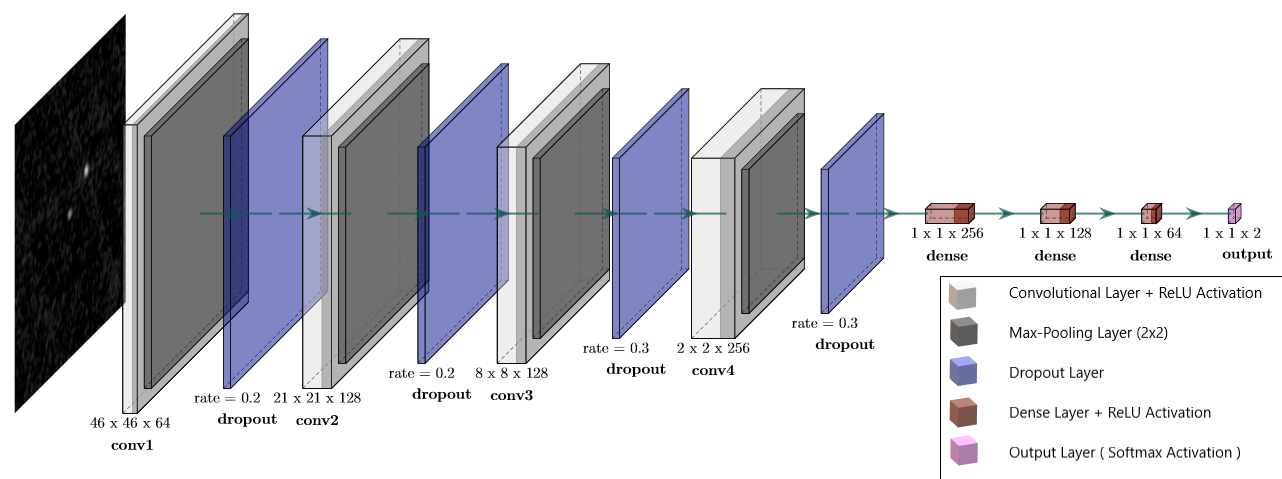}\\
	\includegraphics[scale=0.4]{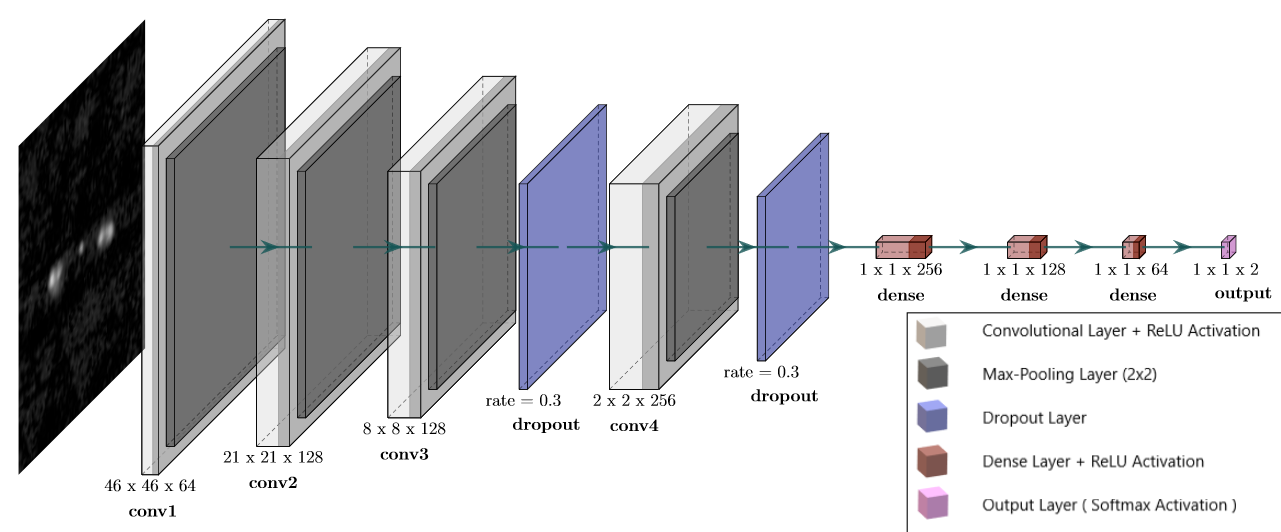}\\
	\end{tabular}

    \caption{The CNN architecture for the model used for the (a)  broad classification (top panel), (b) sub-classification of double sources (middle panel), and (c) sub-classification of triple sources (bottom panel).  }
    \label{fig:cnn1}
\end{figure*}
As the test accuracy for a single CNN model was not good enough,  we followed an approach to first design a CNN model that broadly classified the sources into core, doubles, triples, and jets. We then perform the sub-classification for 'double' and 'triple' sources using two separate models. 
 The sources classified as 'doubles' and 'triples' using the first broad classification are used as inputs for the sub-classification CNN models. This approach is helpful to reduce the misclassification caused due to the similarity between double and triple lobe sources (for e.g., between a double and an irregular triple with one bright and one faint lobe). It also allows us to independently design architectures for feature-specific classification (i.e., number of extended emissions or core/lobe dominance).

The input images are grey-scaled images of 50 x 50 pixels. The number of core, double and triple  sources included in the training set of the broad classifier can be found in Table.~\ref{tab:imgaug}. We experimented with varying the number of layers, the number of neurons, and the size of convolution kernels to arrive at a structure with three 2-D convolutional layers having 64, 128, and 128 kernels of size 5x5, 3x3, and 3x3 pixels, respectively. The outputs from each convolutional layer have a 2x2 max pooling layer applied to reduce dimensionality. A couple of dropout layers are applied to reduce over-fitting. We set the dropout rate at 0.25 and 0.2 for the first and second dropout layer seen in Fig.~\ref{fig:cnn1}. We used the Rectified Linear Unit (ReLU) activation function for each of the convolutional layers. After the third convolutional layer, we used two one-dimensional dense layers of fully connected 128 and 64 neurons, respectively, with ReLU activation. The final layer is a dense layer with 4 neurons corresponding to 4 morphological classifications. This layer is assigned a softmax activation function to obtain a categorical classification of data.

For the sub-classification of the double lobes into core-dominated and lobe-dominated doubles, we followed a CNN structure with four convolutional layers with 64, 128, 128, 256 kernels of size 5x5, 3x3, 3x3, and 3x3 pixels, respectively. We applied a 2x2 max pooling after each convolutional layer. A rate of 0.2 is assigned to the first two drop-out layers, and a rate of 0.3 is assigned to the last two drop-out layers. Three dense layers of 256, 128, and 64 neurons are placed after the fourth convolutional layer. All the layers mentioned above are assigned with the ReLU activation function. The final layer is a dense layer with 2 neurons corresponding to the 2 classification labels.
The architecture for the sub-classification of triple lobes into core-dominated and lobe-dominated triples is similar to that of the doubles, except that the first two convolutional layers are not followed with any drop-out layers.

In all the three CNN models, we used sparse categorical cross-entropy to determine the loss of our network and the "adam" adaptive learning rate optimization algorithm \citep{Kingma2014}.
The architecture of the three models are shown in Fig.~\ref{fig:cnn1}.

We applied a 25 \% validation split on the training data for the broad classifier and a 30 \% split for the sub-classifier models. We trained the models for 15 epochs with a batch size of 100 for the broad classifier and double's sub-classifier and a batch size of 80 for the triple's sub-classifier. Fifteen epochs were deemed good enough for the models as we saw a rise in the validation loss afterward. The confusion matrix for the validation set and the learning curve of these models are shown in Fig. \ref{fig:cnnacc} in the appendix.

\subsection{CNN classifier results}
\label{subsec:morphclass}
\begin{table*}
	\centering
	\caption{Precision, Recall and F1 score for all the classes under the three CNN models discussed in section \ref{sec:cnn}.}
	\label{tab:performance}
	
	  \begin{tabular}{lccccr} 
		\hline
		Model & Class & Validation Sample (Augmented) & Precision & Recall  & F1 Score\\
		\hline
		Main & Core & 2447 & 0.9908 & 0.9779 & 0.9846\\
		 & Triple & 2496 & 0.9935 & 0.9823 & 0.9878\\
		 & Double & 2452 & 0.9823 & 0.9959 & 0.9889\\
		 & Jet & 2109 & 0.9765 & 0.9826 & 0.9824\\ \\
		Double & Core-dominated & 2712 & 0.9987 & 0.9957 & 0.9971\\
		 & Lobe-dominated & 2351 & 0.9963 & 0.9988 & 0.9975\\ \\
		Triple & Core-dominated & 2452 & 0.9938 & 0.9938 & 0.9937\\
		 & Lobe-dominated & 2508 & 0.9940 & 0.9940 & 0.9939\\
		\hline
	   \end{tabular}
    
\end{table*}
In order the evaluate the performance of our CNN models, we make use of the F1 score. The F1 score is an evaluation metric for classification algorithms that uses precision and recall for a specific class. It is the weighted average of precision and recall with a high F1 score of 1 showing an ideal classification and a low F1 score of 0 expressing a failed classification. The F1 score is expressed as 

\begin{equation}
    \text{F1} = 2  \left(\frac{\text{Precision $\times$ Recall}}{\text{Precision + Recall}}\right)
	\label{eq:f1score}
\end{equation}

where the precision is a measure of correctly classified samples, and the recall is a measure of the sensitivity of the classifier model. Equations \ref{eq:precision}  are used for calculating the precision and recall respectively. 
\begin{equation}
    \text{Precision} = \dfrac{\text{TP}}{\text{TP + FP}};\hspace{1cm}   \text{Recall} = \dfrac{\text{TP}}{\text{TP + FN}}
	\label{eq:precision}
\end{equation}

TP and FP stand for true positives and false positives, respectively, while FN stands for false negatives. For example, if we look at triples, true positives are the sources for which the actual and predicted labels are the same. We count it as a false positive for sources predicted as triples but with a different actual label. Similarly, for the actual label of triples, predicted otherwise sources are counted as false negatives. 

Table \ref{tab:performance} shows the precision, recall, and F1 score for all the classes under each of the CNN models. F1 score is the harmonic mean of precision and recall values. The average score for each model is also mentioned in the table. We can see that all three models show high precision, recall, and F1 scores. The high F1 scores indicate that only a tiny fraction of the validation sample is misclassified.

Using the CNN models mentioned before, we classified our radio quasar sample into different morphological classes. The number of sources in each morphological subclass is compiled in Table~\ref{tab:classified_agn}. In this effort, we have increased the sample size of core and doubles by four times, jets by three times, and triples by two times compared to our training set's sample size. 

The "core" type sources can be divided into two sub-classes, based on whether they are resolved or not in the FIRST survey. The "resolved" core sources are expected to consist of quasars with very young radio emission and/or  at a high redshift where the lobes appear as a single component. The "unresolved" sources are expected to have boosted radio cores without any extended radio emission, or it may be a distant quasar with an unresolved extended emission. A dimensionless concentration parameter (Equation \ref{eq:core_res}) is defined in \cite{Kimball2011}, as the ratio of integrated flux density (S\textsubscript{20}) to the peak flux density (S\textsubscript{peak}) at 20 cm.

\begin{equation}
    \theta =\sqrt{\frac{S\textsubscript{20}}{S\textsubscript{peak}}}
	\label{eq:core_res}
\end{equation}

The "resolved" core sources have log($\theta^{2}$) > 0.05 and "unresolved" core sources have log($\theta^{2}$) $\leq$ 0.05. For our further analysis of BAL quasar orientation, we mainly focused on the resolved core quasars as  the exact morphology of unresolved core sources is unknown. 

\begin{table}
	\centering
	\caption{The table lists the number of sources in each morphological class. The first column refers to the main class of the quasar source, while the second column indicates the sub-classification for the source. The number of sources in each sub-class is mentioned in the third column.}
	\label{tab:classified_agn}
	\begin{tabular}{l | c | r} 
		\hline
		Class & Sub-class & No. of Sources\\
		\hline
		Core & Resolved & 10441\\
		Core & Unresolved & 2526\\ \\
		Double & Core-dominated & 1246\\
		Double & Lobe-dominated & 548\\ \\
		Triple & Core-dominated & 254\\
		Triple & Lobe-dominated & 926\\ \\
		Jet & - & 550\\
		\hline
	\end{tabular}
\end{table}
\section{Orientation of BAL quasars}
\label{sec:balquasars}
It is believed that the difference between the core only, core-dominated, and lobe-dominated sources arise mainly due to the difference in the orientation angle with respect to the line-of-sight \citep{Orr1982, Kapahi1982}. If orientation is the main factor responsible for the BAL phenomena, we expect a difference in the BAL fraction between these different radio morphological classes.

\subsection{BAL fraction in different morphological classes}
\label{sec:balfrac}

\begin{table}
	\centering
	\caption{The total number of BALs and non-BALs in each morphological class is listed below. The calculated BAL fraction values are listed in the last column.{ The error in BAL fraction is computed using Poisonian statistics.}}
	\label{tab:balf}
	\begin{tabular}{l | c | c | r} 
		\hline
		Class & No. of BALs & No. of non-BALs & BAL fraction\\
		\hline
		Resolved Core & 286 & 976 & 0.23 $\pm$ 0.01\\
		Core-dominated & 165 & 499 & 0.25 $\pm$ 0.02\\
		Lobe-dominated & 107 & 356 & 0.23 $\pm$ 0.03\\
		\hline
	\end{tabular}
\end{table}

In order to obtain the number of BALs and non-BALs from our sample of 16491 sources, we used the BAL$\_$PROB parameter from the DR16Q catalog.  We classified the sources with BAL probability (BAL$\_$PROB) > 0.5 as BAL quasars and the rest as non-BAL quasars. DR16Q also has quasars BAL$\_$PROB = -1  which are the low redshift sources whose spectra did not cover the \civ\ and \siiv\ wavelengths. We did not include 7697 sources with BAL$\_$PROB = -1 in our analysis as the BAL nature of these sources is uncertain. The resulting sample has 2182 BAL quasars and 6612 non-BAL quasars.   
From hereon, we combined the core-dominated doubles and core-dominated triples into a single class called core-dominated quasars (hereafter, CDQ). Similarly, the lobe-dominated doubles and lobe-dominated triples are combined into a single class, namely lobe-dominated quasars (hereafter, LDQ). Apart from LDQs and CDQs, we also considered resolved core quasars (hereafter, RCQ) in this analysis. We obtained a total of 664 CDQs(BALs: 165,  non-BALs: 499), 463 LDQs(BALs: 107, non-BALs: 356) and 1262 RCQs.

\begin{figure}
	\includegraphics[width=\columnwidth]{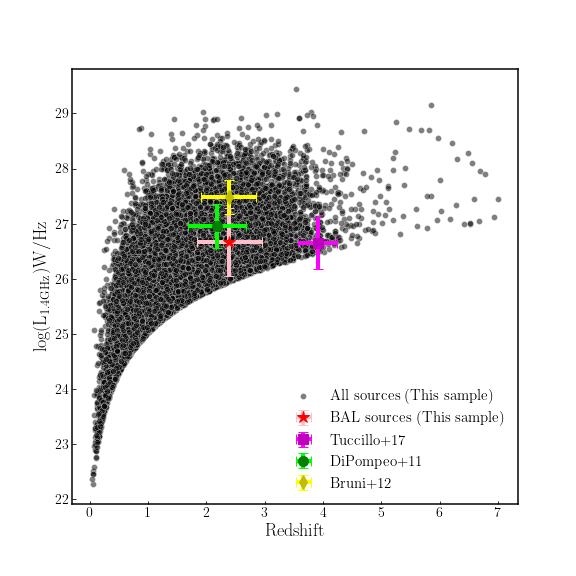}
    \caption{The figure shows the distribution of radio luminosity of the sources in our sample as a function of redshift. The loci of the BAL quasars in our study and from samples of previous studies have been marked. The star, square, circle and diamond points correspond to the median values of BAL sources in our sample, \citet{Tuccillo2017}, \citet{ Bruni2012}, and \citet{DiPompeo2011a} samples.   The error bars correspond to  1-$\sigma$ of the values for each of the samples. }
    \label{fig:balcut}
\end{figure}
{  Fig~\ref{fig:balcut} shows the radio luminosity vs redshift comparison of our sample with previous BAL quasar samples \citep{Tuccillo2017, Bruni2012, DiPompeo2011a}. The sample in \cite{Tuccillo2017} consists of 22 BAL sources with 3.6 $\leq$ z $\leq$ 4.8, detected both in SDSS DR7 and VLA FIRST surveys, and having radio loudness parameter greater than 10. Sample provided by \cite{Bruni2012} comprises 25 BAL quasars found in the SDSS DR4 catalog with a counterpart in VLA FIRST, having S$_{1.4}$ $>$ 30 mJy. \cite{DiPompeo2011a} catalog consists of 74 BAL quasars that are common in FIRST and NVSS surveys with integrated flux density greater than 10 mJy and z $\geq$ 1.5. Unlike our sample, the previous studies have mainly focused on BAL quasars with high radio fluxes. The median radio luminosity of our BAL quasar sample is lower than the \citet{Bruni2012} and \citet{DiPompeo2011a} samples and comparable to the \citet{Tuccillo2017} sample.  Though the \citet{Tuccillo2017} sample is at high redshift, the median redshifts of the \citet{Bruni2012} and \citet{DiPompeo2011a} samples are comparable to our sample.} 

The number of BALs and non-BALs in three different morphological classes is shown in Table \ref{tab:balf} along with the corresponding BAL fraction values. Resolved core, core-dominated, and lobe-dominated sources have a BAL fraction of 0.23 $\pm$ 0.01, 0.25 $\pm$ 0.02, and 0.23 $\pm$ 0.03 respectively. We do not see any difference in the BAL fraction between the core only, core-dominated, and lobe-dominated quasars and all the three morphological classes have a BAL fraction close to 23\%.  This is consistent with the previous studies that have found intrinsic fractions closer to 20 percent or higher from the infrared and radio wavebands.  A detailed discussion on the observed BAL fractions is given in section.~\ref{sec:discussion}.

The difference in sample sizes of different morphological classes can introduce some bias in the BAL fraction analysis.
To remove any such biases, we computed the BAL fraction in each morphological class by randomly selecting sources with a sample size defined by the smallest sample size among the three morphological classes. Lobe-dominated sources have the smallest sample size of 463 sources. We randomly selected a set of 463 sources from each of the other two classes and measured the BAL fractions in that respective sets. We repeated this procedure for 1000 iterations by randomly shuffling the sources within a morphological class for each iteration.  The median value of the BAL fraction distribution for a morphological class is taken as the unbiased BAL fraction for that class. This exercise resulted in BAL fractions of 0.23 $\pm$ 0.02 and 0.25 $\pm$ 0.01  for the resolved core and core-dominated sources, respectively.

 We observe that the BAL fraction for all three classes is nearly the same.  If the distribution of the number of sources as a function of orientation is different for the different classes, it may bias the BAL fraction measurements.  The average BAL fraction is mainly driven by the bins having the most significant number of sources. If those bins in the three classes, albeit at different orientation angles, somehow have the same BAL fractions, we will not see any difference in the BAL fraction between the three different morphological classes. To probe further into this, we  studied the BAL fraction as a function of some quasar orientation indicators. 

\subsection{Quasar orientation indicators}

Two common parameters used to study the orientation of quasars are the core-to-lobe flux density ratio (R) and core radio loudness parameter (R$_I$). The definition for the R and R$_I$ parameters in this paper have been adopted from \cite{Kimball2011}.

\begin{equation}
     R = \dfrac{S\textsubscript{core}}{S\textsubscript{lobe}} (1+z)^{(\alpha \textsubscript{lobe} - \alpha \textsubscript{core})},
	\label{eq:Rpar}
\end{equation}

\smallskip
where S\textsubscript{core} and S\textsubscript{lobe} refer to the 20 cm flux densities of the core and lobes respectively. We set the values for core spectral index $\alpha$\textsubscript{core} = 0 and lobe spectral index $\alpha$\textsubscript{lobe} = -0.8.

\begin{equation}
     log(R_I) = \dfrac{L\textsubscript{radio}}{L\textsubscript{optical}} = \dfrac{M\textsubscript{radio} - M\textsubscript{i}}{-2.5},
	\label{eq:R1par}
\end{equation}
\smallskip
where M\textsubscript{radio} is the K-corrected radio absolute magnitude at 20 cm \citep[refer Eqn.5 of][for the exact definition of  M\textsubscript{radio}]{Kimball2011}   and M\textsubscript{i} is the Galactic reddening corrected and K-corrected absolute i-band magnitude. The M\textsubscript{i} values are taken from the DR16Q catalog.  We exclude 85 sources that do not have a valid M$_i$ value. This results in a dataset with 659 CDQs ( BALs: 162,  non-BALs: 497), 462 LDQs ( BALs: 107,  non-BALs: 355), and 1260 RCQs.

To measure the core-to-lobe flux ratio (R), we need the integrated flux density (S\textsubscript{20}) of the core and lobe separately. To measure these flux density values, we performed a radial search for radio emissions detected by VLA FIRST within 4' around the core of objects classified as 'doubles' or 'triples' by our CNN model. For radio emissions found around the objects classified as 'doubles', we set a limit of one emission value nearest to the core, and for the objects classified as 'triples', we set a limit of two radio emissions nearest to the core. For 'triples', the lobe with the highest flux is used to get the S\textsubscript{lobe} value. It should be noted that the R value measured using this technique might not provide the correct value for sources where multiple radio emissions are found between the core and lobe.  The number of BALs and non-BALs with a corresponding R value include: 637 CDQs (BALs: 157,  non-BALs: 480) and 456 LDQs (BALs: 106,  non-BALs: 350). There are some sources included in the R$_I$ dataset for which we could not identify the lobe component in VLA FIRST using our procedure. In order to have a similar dataset while analyzing R and $R_I$ parameters, we use the 632 CDQs (BALs: 154,  non-BALs: 478) and 455 LDQs(BALs: 106, non-BALs: 349) that are common in both datasets.

\begin{figure}

	\includegraphics[width=\columnwidth]{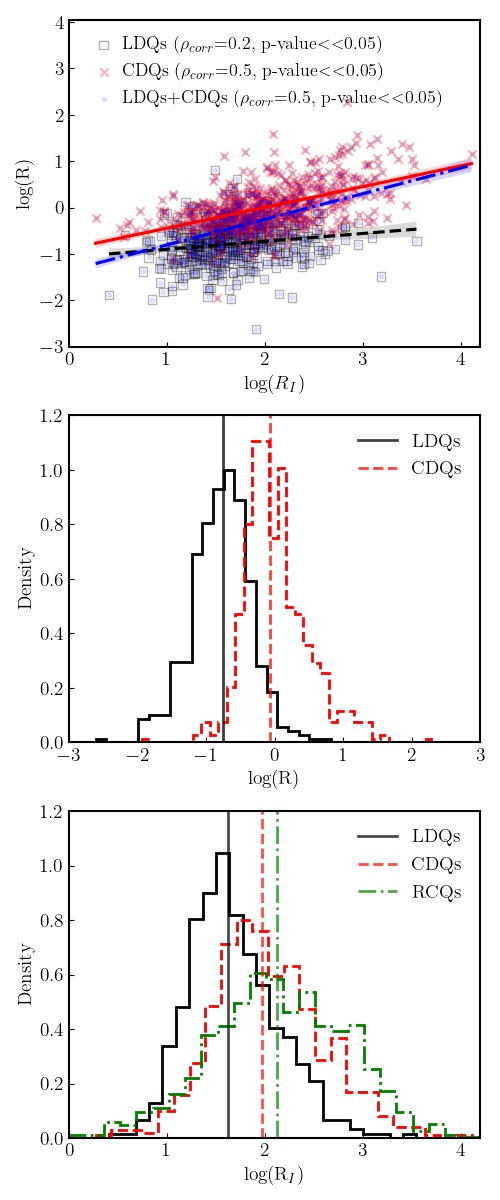}
    \caption{Top panel: Spearman-rank correlation analysis between the two quasar orientation indicators, log(R) and log(R$_I$) for lobe-dominated quasars (LDQs), core-dominated quasars (CDQs), and a combined sample of LDQs+CDQs. The dashed/black, solid/red, and dot-dashed/blue lines represent the best fit for the LDQs, CDQs, and the combined sample.  Correlation coefficients and p-values are given in the legend.  Middle panel: The distribution of log(R) for LDQs (solid/black) and CDQs (dashed/red).   Bottom panel: The distribution of log(R$_I$) for LDQs (solid/black), CDQs (dashed/red), and resolved core quasars (RCQs, dot-dashed/green). The  vertical lines show the median values for the corresponding distributions.}
    \label{fig:log(R)vsR1}
\end{figure}

As discussed in section \ref{sec:intro}, several studies have pointed out that R$_I$ is a better statistical indicator of orientation than R. Both these indicators are measures of the radio core boosting factor. The advantage of R$_I$ over R mainly arises from the normalization used in the core boosting factor. R  normalizes using the extended lobe emission flux, which is subjected to aging and environmental effects. Contrarily, R$_I$ normalizes using core optical flux, independent of the age and environment. To better understand these effects,  we first checked for any correlation between the two quasar orientation indicators defined above. Although with a large scatter, \citep{Kimball2011} has shown that there is indeed a correlation between these parameters in their sample. They also argued that the observed correlation is not a selection effect but is an intrinsic property. The top panel of Fig.~\ref{fig:log(R)vsR1} shows the correlation between R and  R$_I$ for core-dominated and lobe-dominated sources. We see that core-dominated quasars have a moderate correlation with a spearman-rank correlation coefficient of 0.49 and a p-value of 2.45$\times$10$^{-40}$. Lobe-dominated quasars are weakly correlated with a spearman-rank correlation coefficient of 0.20 and a p-value of 1.2$\times$10$^{-5}$.   Including both the samples result in a correlation coefficient of 0.46 and a p-value of 3.7$\times$10$^{-59}$.  We also see a large scatter in the correlation suggesting that other factors like age, environment are also influencing these parameters. A weaker correlation for the LDQs may suggest that LDQs may be more severely affected by these factors than the CDQs.

 Middle panel of Fig.~\ref{fig:log(R)vsR1} shows the distribution of log(R) for CDQs and LDQs. The definition of CDQ and LDQ itself points to a  clear separation in log(R) distribution between the two classes. Indeed, we also see that the distributions of log(R) for CDQs and LDQs are well separated.   The bottom panel of Fig.~\ref{fig:log(R)vsR1} shows the log(R$_I$) distributions of CDQs and LDQs together with the  RCQs. The  vertical lines represent the median of the corresponding distribution. Here again, although we see a progression in the median of log(R$_I$) distribution for LDQs, CDQs, and RCQs, there is significant overlap between these distributions. It can be seen that the CDQs appear to have a higher median log(R$_I$)  than the LDQs. The distribution for RCQs is shifted to even higher log(R$_I$) values than CDQs. Kolmogorov–Smirnov (KS) test between the different samples results in statistically significant p-values ($<<$0.05), indicating that the different samples are not drawn from the same distribution.

\begin{figure*}
    \centering
    \includegraphics[scale=0.5]{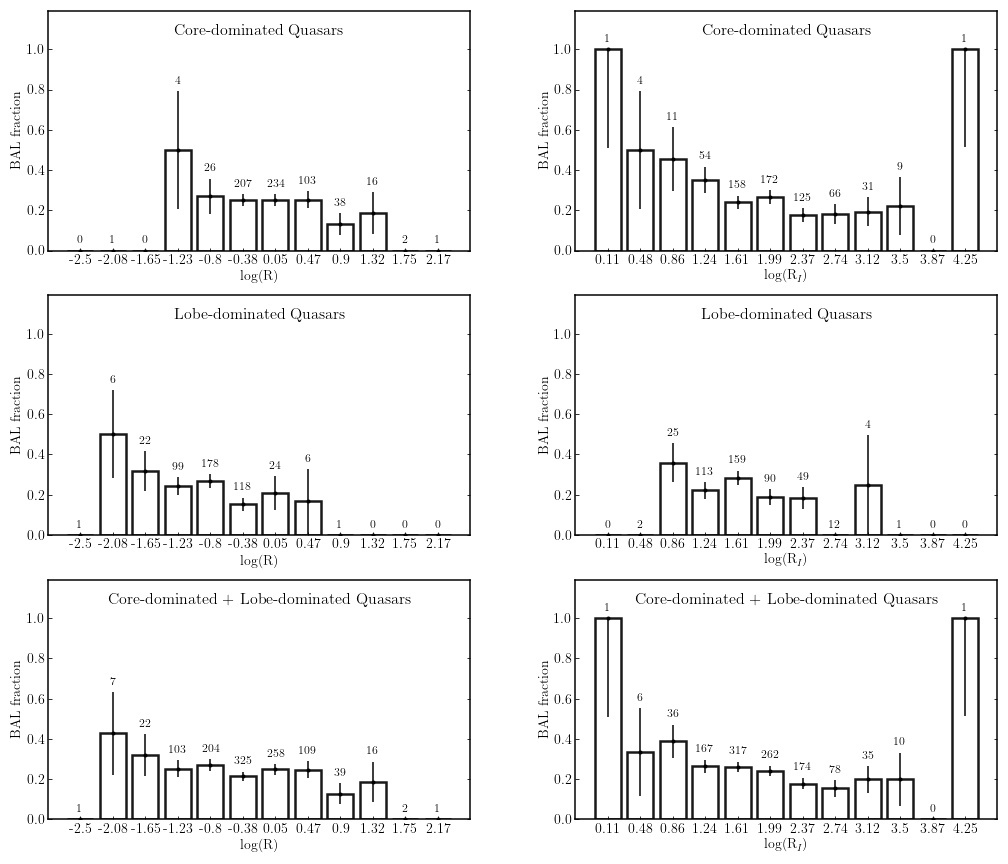}
    \caption{BAL fraction vs. log(R) and log(R$_I$) for core-dominated quasars (CDQs) and lobe-dominated quasars (LDQs). The left column plots show the variation of BAL fraction as a function of orientation indicator, log(R) for CDQs, LDQs, and a combined sample of CDQs and LDQs. The right column plots show the variation of BAL fraction as a function of another orientation indicator, log(R$_I$) for CDQs, LDQs, and a combined sample of CDQs and LDQs. The error bars in each log(R)/log(R$_I$) bin are  one sigma bootstrap errors. The number of sources in each bin is marked above the bin.}
    \label{fig:log(R)cl}
\end{figure*}

\subsection{Dependence of BAL fraction on quasar orientation}

Having defined two metrics for measuring the orientation of quasars, we studied the distribution of BAL fraction as a function of the orientation indicators. We first looked into the BAL fraction in bins of log(R) and log(R$_I$) for different morphological classes. Fig.~\ref{fig:log(R)cl} shows the BAL fraction in different log(R) (left column panels) and log(R$_I$) (right column panels) for LDQs (top row panels), CDQs (middle row panels) and a combined LDQ+CDQ sample (bottom row panels). The error bars represent one sigma bootstrap error where we re-sampled  1000 times and computed the BAL fraction in each bins for each iteration. The number of sources in each bin is also marked above the bin. The core-boosting parameters R and R$_I$ are higher for small orientation angles of the jet with respect to the line of sight.  From Fig.~\ref{fig:log(R)cl}, it is clear that the BAL fraction almost doubles for a low value of orientation indicators (i.e., high orientation angles) as compared to high values. In the case of the R parameter, the increase of BAL fraction towards high orientation angles is evident for both CDQs and LDQs. As R is defined as the ratio of core-to-lobe flux, it is not surprising that the log(R) distribution of LDQs is shifted to low R values compared to the CDQs. However, such a clear separation of  R$_I$ parameter distribution is not seen between the CDQs and LDQs. In fact, some of the CDQs have larger orientation angles than the maximum orientation angle of LDQs. For log(R$_I$) distribution, the increase in BAL fraction for high orientation angles is very prominent for the CDQs. The trend is less prominent for LDQs as the R$_I$ values are more clustered into two or three bins. The small number of sources may heavily bias the highest orientation angle bins. To improve the statistics, we also looked at the distribution of log(R) and log(R$_I$) with a combined sample of CDQs and LDQs. The bottom left and right panels of Fig.~\ref{fig:log(R)cl} show the variation of BAL fraction as a function  of log(R) and log(R$_I$), respectively, for the combined sample. It is clear from these plots that the BAL fraction increases from 0.2 to 0.4 with the increase in the orientation angle. 

\begin{figure*}
    \centering
    \includegraphics[scale=0.5]{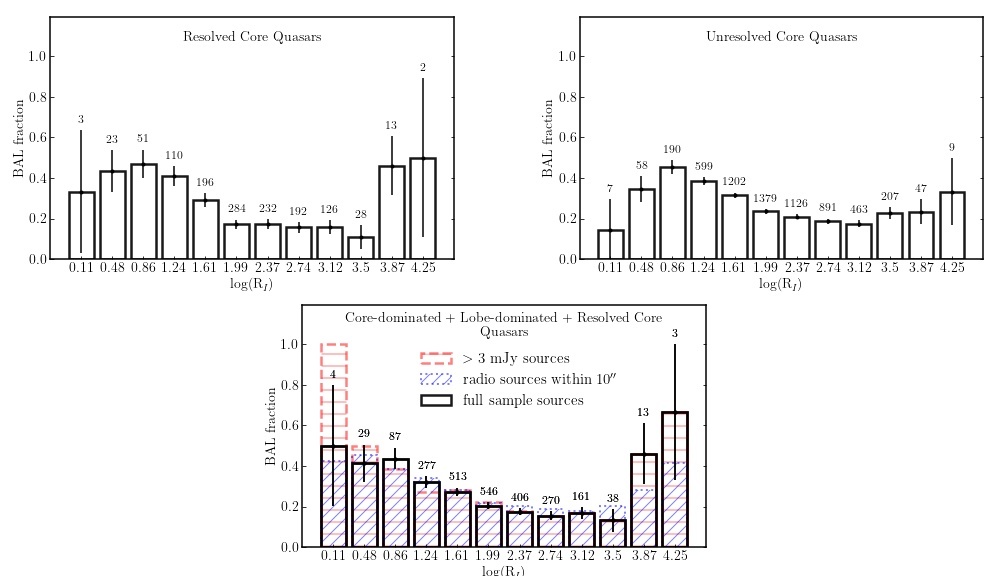}
    \caption{BAL fraction vs. log(R$_I$) for resolved cores, unresolved cores, and a combined sample of core-dominated, lobe-dominated, and resolved core quasars. In all the panels, the error bars represent one sigma bootstrap error. The total number of sources in each bin is marked above the bin. The top two panels show the variation of BAL fraction with log(R$_I$) for resolved and unresolved core quasars. The bottom panel shows the BAL fraction vs. log(R$_I$) trend for a combined sample of core-dominated, lobe-dominated, and resolved core quasars. The open/black, horizontally hatched /red, and 45 degree hatched/blue histograms are for full sample sources, sample with sources having FIRST flux greater than 3 mJy, and sample whose radio counterparts are within 10'' of the optical emission, respectively.  }
    \label{fig:log(R)cl2}
\end{figure*}
Since the R$_I$ parameter depends only on the radio and optical flux of the core, we can use it to study the distribution of BAL fraction within the morphological class of 'core only' sources with no lobe emission.
 The top two panels of Fig.~\ref{fig:log(R)cl2} shows the distribution of log(R$_I$) parameter for resolved (left panel) and unresolved core (right panel) only quasars. Here again, it is clear that the BAL fraction increases for higher orientation angles. Unresolved core quasars may consist of either sources with boosted cores with undetectable extended radio emissions or sources whose extended radio emission is not resolved in the VLA FIRST survey. Hence, the behavior of the R$_I$ distribution for unresolved cores can be a sum of distributions from different morphologies. The open/black histogram in the bottom  panel of  Fig.~\ref{fig:log(R)cl2} shows the log(R$_I$) distribution for a combined sample of RCQs, CDQs, and LDQs, which again confirms the BAL fraction trend. 
 
 Our sample may suffer from selection effects due to the incompleteness of the FIRST survey. The completeness of the FIRST survey is about 93\% at 3 mJy, which falls to 53 \% at 1.1 mJy \citep{Jiang2007}. To examine if our results are affected by this incompleteness, we compared the trend of BAL fraction increase at high orientation angles for two samples; one with the full sample and another with a sample defined by only those radio sources with FIRST integrated flux, f$_{int}$ $>$ 3 mJy (horizontally hatched/ red histogram in Fig.~\ref{fig:log(R)cl2}). Even with this cut, the sharp increase in the BAL fraction towards high orientation angles remains unchanged. Another bias that can affect our results is due to incomplete radio identifications in the DR16Q sources. We initially adopted a search radius of 2'' within the optical emission to identify the radio counterpart.  To ensure that the chosen search radius does not introduce any bias, we generated a new radio quasar sample by cross-matching the SDSS and FIRST with a relaxed search radius of 10'' within the optical source. The  45 degree hatched/blue histogram in the bottom panel of Fig.~\ref{fig:log(R)cl2}) shows the BAL fraction with the radio counterpart defined by the nearest source in FIRST within 10''.   We also tested for biases from redshift or luminosity selection effects by repeating the analysis using sources limited to specific ranges of redshift and luminosity. In all these checks, we see an increase in the BAL fraction at high orientation angles which rule out any biases in our analysis.
 
 Using radio variability arguments, several studies have shown that some BAL quasars are viewed along the polar axis \citep{Zhou2006, Ghosh2007}. As the number of these polar BAL quasars is small, it is not yet clear if they constitute an entirely different class of outflows. There also have been some criticisms in the literature questioning the polar nature of these sources, i.e., whether the radio variability can be attributed to beaming effects or not \citep{Hall2011}. \citet{Borguet2010} simulated the two-component polar+equatorial wind model of quasar outflows and showed that it is necessary to include both the equatorial and the polar absorption regions to explain the \civ\ BAL profiles in some sources. However, the search of \citet{Mishra2019} to find BAL blazars using intra-night optical variability did not result in any detection.
Interestingly, we find that the BAL fraction increases for the highest log(R$_I$) (lowest orientation angles) bins even though the number of sources in these bins is low.  If confirmed with a larger sample, this BAL fraction increase at low orientation angles may be due to the existence of 'polar' BAL quasars.

Fig.~\ref{fig:log(R)cl} \& \ref{fig:log(R)cl2} are also helpful to understand why the average BAL fraction for the three morphological classes are similar (see, section \ref{sec:balfrac}). It is clear that most sources for the CDQs, LDQs, and RCQs are distributed between the log(R$_I$)  values 1.99 and 3.5. Incidentally, these bins have a BAL fraction $\sim$ 20 \% and these bins dominate the computation of the average BAL fraction.

\subsection{LoBAL and FeLoBAL quasars}
\label{sec:hilo}
In the orientation model of BAL quasars, LoBALs and FeLoBALs are thought to be arising when the line of sight is closer to the equatorial accretion disk. As most of the true edge-on quasars would be obscured by the dusty torus, what we mean here is that they are seen from a viewing angle farther from the accretion disk symmetry axis compared to non-BAL quasars.  In this model, LoBALs and FeLoBALs are more reddened than HiBALs as the column density of the absorbing material is high when the line of sight is closer to the accretion disk. Hence, we expect an increase in the BAL fraction at high orientation angles for LoBALs and FeLoBALs. We investigated this phenomenon using a sample of sources classified as HiBALs, LoBALs, and FeLoBALs in the  \cite{Trump2006} catalog. We checked the distribution of the BAL fraction in different log(R$_I$) bins for the two BAL quasar sub-classifications.

Since the BAL quasars in \cite{Trump2006} are taken from the SDSS DR3Q catalog, we first cross-matched the SDSS DR3Q catalog with the VLA FIRST survey to obtain a sample of BAL and non-BAL quasars. Our sample consists of 3893 sources in total, among which 477 are BAL quasars. Using the different BAL sub-type flag values in the \citet{Trump2006} catalog, we identified  335 HiBALs, 100 LoBALs and 42 FeLoBAL sources.

\begin{figure}
	\includegraphics[width=\columnwidth]{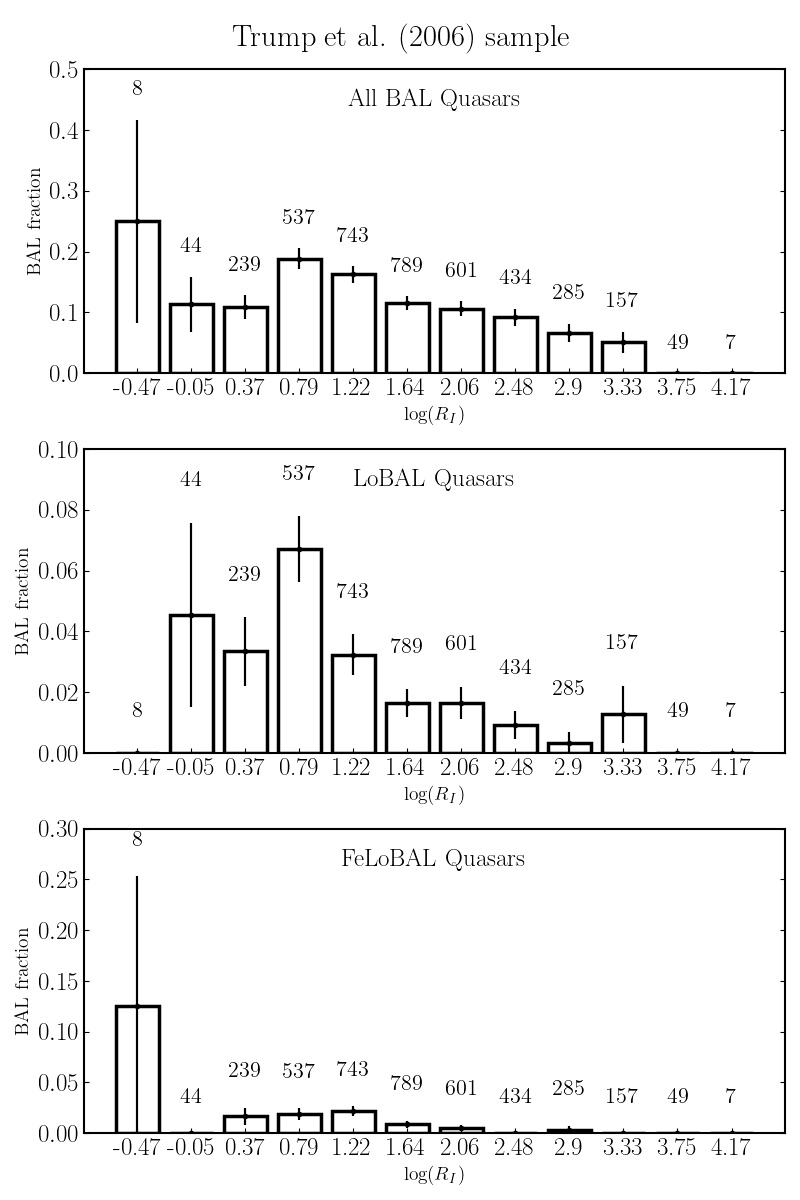}
    \caption{BAL fraction vs log(R$_I$) distribution for All BAL, LoBAL, and FeLoBAL quasars from the \citet{Trump2006} catalog. In all the panels, the error bars in each log(R$_I$) bin are  one sigma bootstrap errors. The total number of sources in each bin is marked above the bin.}
    \label{fig:hilobal}
\end{figure}
The top panel in Fig.~\ref{fig:hilobal} shows the log$R_I$ distribution for total BAL fraction (HiBALs + LoBALs + FeLoBALs) in our sample of 3893 quasars. We observe that the distribution in the top panel is similar to the result that we obtained for our DR16Q sample.  This confirms the trend of BAL quasar fraction increase for high orientation angles using an independent dataset. The middle and bottom panels in Fig.~\ref{fig:hilobal} show the log$R_I$ distribution for LoBALs and FeLoBALs respectively. LoBALs and FeLoBALs also reaffirm the previously seen trend of BAL fraction increase at high orientation angles. To quantify the amount of BAL fraction increase, we calculated the percentage of BAL fraction increase between the two log($R_I$) bins in the upper and lower log(R$_I$) values,  which at least have more than 50 sources. We see that the percentage of BAL fraction increases by 53 \%, 163 \%, 376 \%, and 113 \% for the HiBAL, LoBAL, FeLoBAL, and total samples, respectively, with the increase in the orientation angle. Clearly, the rate of increase of BAL fraction is maximum for  FeLoBALs, intermediate for LoBALs, and minimum for HiBALs. The progressive reduction in the rate of BAL fraction change among FeLoBALs, LoBALs, and HiBALs adds further evidence to the orientation model of BAL quasars where FeLoBALs and LoBALs are preferentially seen at higher orientation angles as compared to HiBALs. 

\begin{figure}
	\includegraphics[width=\columnwidth]{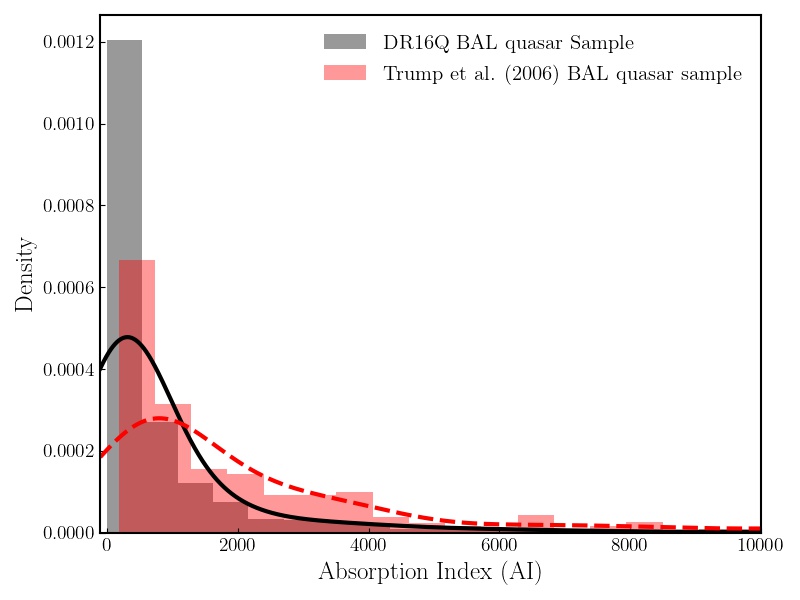}
    \caption{Comparison of \civ\ absorption index for BAL quasar sources in the DR16Q  and \citet{Trump2006}   catalog. The grey histogram corresponds to the DR16Q catalog, and the red histogram corresponds to the \citet{Trump2006} catalog. The corresponding KDE plots are also shown as solid/black and dashed/red lines for DR16Q and \citet{Trump2006} catalog, respectively. }
    \label{fig:AI}
\end{figure}

Interestingly, we also note that the BAL fractions computed for \citet{Trump2006} catalog sources (Fig.~\ref{fig:hilobal}) are considerably lower than the BAL fractions computed for DR16Q catalog sources (Fig.~\ref{fig:log(R)cl} \& \ref{fig:log(R)cl2}). We probed the difference between these two datasets to understand the reduction in the BAL fraction. We first compared the radio core intensity and absolute I-mag distributions of the sources in the  \citet{Trump2006} catalog and DR16Q catalog that are included in our analysis. We do not see any difference in the distribution of the FIRST radio intensity between these samples. However, we see a notable difference in the absolute I-mag distribution with the \citet{Trump2006} catalog sources shifted towards the brighter side by two magnitudes. This is expected as the   SDSS DR3, being one of the earliest SDSS data releases, targeted relatively brighter quasars. However, it is not clear whether this difference in the optical luminosity affects the BAL fraction. A more prominent and clear difference is obtained in the distribution of absorption index (AI) between the \citet{Trump2006} and DR16Q catalog sources. The absorption index is a metric, first proposed by \citet{Hall2002}, to characterize the presence of BAL features in quasar spectra. It is a measure of equivalent width, measuring all absorption within the limits of every trough when the integration limit extends from 0 to 29,000 \kms. Fig.~\ref{fig:AI} shows the distribution of AI measured for \civ\ region for the \citet{Trump2006} and DR16Q catalog sources. It is clear that the DR16Q catalog has much more sources with small AI values as compared to the \citet{Trump2006} catalog sources. This difference in AI  can be attributed to the difference in the continuum normalization procedure and the BAL trough definition between the two catalogues. As DR16Q employed a PCA-based continuum normalization and an automated BAL identification procedure using a CNN, it should be sensitive to even low values of AI. It is possible that \citet{Trump2006} et al. have adopted a more stringent criteria for the identification of BAL troughs in quasars.

\section{Discussion}
\label{sec:discussion}
\begin{figure}
	\includegraphics[width=\columnwidth]{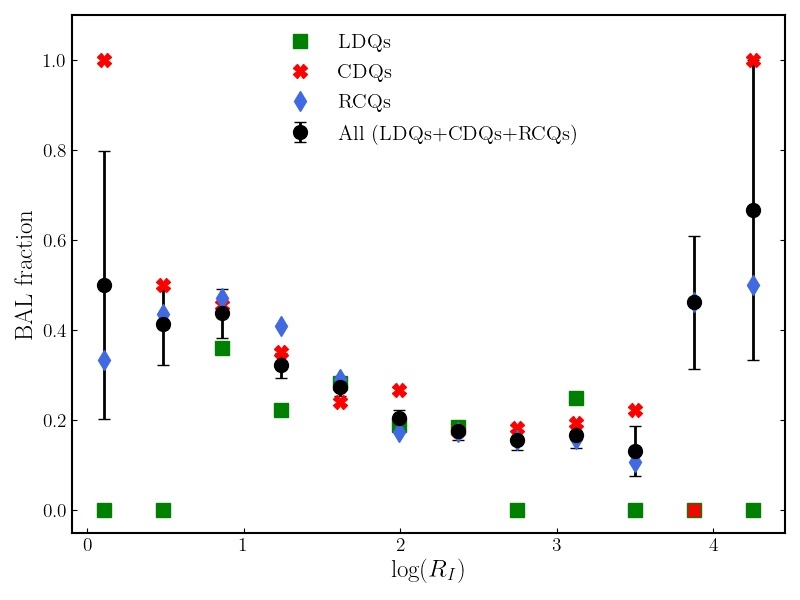}
    \caption{BAL fraction vs. log(R$_I$) for the resolved core quasars (blue diamonds), core-dominated quasars (red cross marks), lobe-dominated quasars (green squares), and a combined sample of resolved core, core-dominated, and lobe-dominated quasars (black circles). The error bars in the BAL fraction indicate the one sigma bootstrap errors  for the combined sample. }
    \label{fig:spall}
\end{figure}

In this study, we probed the orientation of BAL quasars using a large sample of 16491 sources that have spectra available from the SDSS DR16 as well as radio observations available from the VLA FIRST survey. Using CNN models, we first classified this sample into different radio morphological classes such as core only sources, core-dominated lobe sources, and lobe-dominated lobe sources. We found that the average BAL fraction is the same $\sim$ 23 \% for these different morphological classes. Using two quasar orientation indicators, we investigated the BAL fraction distribution as a function of quasar orientation. The main result from this analysis is that the BAL fraction almost doubles at high orientation angles irrespective of the radio morphology. We also repeated this study using samples designed to remove any selection effects and found that the trend of BAL fraction increase at high orientation angles is unchanged. Fig.~\ref{fig:spall} shows the variation of BAL fraction as a function of log(R$_I$) for the resolved core, core-dominated, and lobe-dominated quasars. We also show the trend for a combined sample of resolved core, core-dominated, and lobe-dominated quasars. It is clear that the BAL fraction increases to 0.4 from 0.2  when orientation angle increases (equivalently log(R$_I$)   decreases). A similar trend was also seen for HiBALs, LoBALs, and FeLoBALs using a different sample taken from \citet{Trump2006} catalog. Interestingly, among the three BAL quasar sub-classifications, the rate of change of BAL fraction is maximum in FeLoBALs, intermediate in LoBALs, and minimum in HiBALs.

{ We validated our results against different selection effects due to redshift, luminosity,  incompleteness of the FIRST survey, and incomplete radio counter-part identifications. However, we have not explicitly checked for any bias due to incomplete target selection in the DR16Q catalog. 
As quasars are chosen for spectroscopic observations in SDSS primarily based on their photometric properties, our analysis can be biased due to the differential target selection completeness for  BALQSOs and non-BALQSOs. As BALQSOs  are significantly dust-reddened compared to non-BALQSOs, the completeness of BALQSOs is expected to be less. \citet{Allen2011} computed this difference in the selection probabilities using simulated BAL and non-BALQSO magnitudes and found that the completeness is very high for both  BALQSOs and non-BALQSOs with z $<$ 2.1 and i $<$ 19.1, drops sharply for a narrow region close to z $\sim$ 2.6 and rises again for higher redshifts. This completeness difference can introduce some bias for analysis involving optically selected samples. To our rescue, eBOSS also targets all SDSS point sources that are within 1'' of a radio detection in the  FIRST point source catalog \citep{Myers2015}.  As our final sample is radio-selected BAL quasars and non-BAL quasars, we  do not expect the behavior of the BAL fraction to be severely influenced by any incompleteness effects. As with the case of dust-reddening, strong absorption troughs  can make the broad band magnitude appear fainter for a BALQSO than an equivalent non-BALQSO. \civ\ BAL troughs fall in the SDSS i-band for the redshift range 3.5 $<$ z $<$ 4.4. Our BAL quasar sample contains 82 quasars ($<$ 4 \%) in the redshift range where the i-band luminosities are underestimated due to the presence of \civ\ BAL absorption. Additionally, a small fraction of quasars ($<$ 1 \%) should be LoBALs and FeLoBALs where the i-band luminosities are affected by \mgii, \aliii, and \feii\  broad absorption.  The orientation indicator, radio-to-optical core luminosity  has the i-band optical luminosity in the denominator. Although the fraction of BALQSOs containing strong BAL features within the i-band wavelengths is less , any correction of the optical luminosities  will further push these sources to lower log(R$_I$) bins, thereby increasing the BAL fraction at higher orientation angles. Likewise, \citet{Hewett2003}  suggests that optically bright BAL quasars are half as likely as non-BAL  quasars to be detectable as S$_{1.4 GHZ}$ $\geq$ 1 mJy sources.  To probe this, we divided the sample into three optical luminosity bins, (i) Log(L$_{i-band}$) $\le$ 24 W/Hz, (ii) 24 W/Hz $<$ Log(L$_{i-band}$) $\le$ 25 W/Hz, and (iii) Log(L$_{i-band}$) $>$ 25 W/Hz, and studied the distribution of radio luminosity of BAL and non-BAL quasars. We do not see any difference in the distribution of radio luminosity between BAL and non-BAL quasars for the lower optical luminosity bins. But, for the highest luminosity bin, we see that the BAL quasars have statistically lower radio luminosities as compared to the non-BAL quasars (p-value of KS test $<$ 0.05). Thus, a  flux-limited survey like FIRST would have missed a fraction of optically-bright BAL quasars that have radio fluxes below the FIRST sensitivity. As these missed BAL quasars have low radio luminosities and high optical luminosities, a correction for the missed quasars would only populate the  lower log(R$_I$) bins. This again will  amplify the already seen trend of BAL fraction increase at high orientation angles. To ensure that the optically bright sources  are not biasing the BAL fraction trend, we excluded the optically bright sources from our sample and then studied the variation of BAL fraction as a function of orientation. The pruned sample also follows the trend of high BAL fraction for high orientation angles. }

The fraction of BALQSOs, especially in radio quasars, is an interesting parameter that can constrain the orientation vs. evolution models of BALQSOs. An important issue here is the definition of BALQSO itself. The traditional BALs are defined to have \civ\ absorption troughs broader than 2000 \kms\ and outflow velocities larger than 3000\kms. The fraction of 'traditional' BALQSOs in optically selected quasar samples is typically found to be $\sim$ 10 percent \citep{Weymann1991,Gibson2008}. Although the traditional definition is helpful to include absorption only from nuclear outflows, it also removes a significant fraction of genuine BALQSOs. \citet{Trump2006} redefined the BALQSO definition to include absorption features having narrower widths and lower outflow velocities.  With this new definition,  \citet{Trump2006} found that the BALQSO fraction significantly increases to 26 percent. Studies using infra-red or radio selection have also found the intrinsic fraction to be closer to 20 percent \citep{Dai2008, Ganguly2008, Dai2012,Morabito2014}. Our results for the average BAL fraction in different morphological classes are consistent with the BAL fraction inferred from previous studies. We want to stress here that  BALQSOs in our analysis follows the  \citet{Trump2006} definition rather than the traditional one.

The lack of availability of a reliable indicator of quasar orientation is a major hindrance in quasar studies.
\citet{VanGorkom2015} compared the different indicators of quasar orientation using a sample of 126 radio-loud quasars. Their study mainly compared four indicators that measured radio core dominance as a proxy for quasar orientation. Core dominance compares the radio core luminosity against the intrinsic power of the central engine.  \citet{VanGorkom2015} used four proxies for the intrinsic power of the central engine, namely,  radio lobe luminosity, optical continuum luminosity,  luminosity of the narrow-line region, and  151 MHz luminosity. The main results from their study are : (a) the beamed optical synchrotron emission from the jets is not a significant component of the optical continuum, (b) normalizing the radio core luminosity by the optical continuum luminosity yields a demonstrably superior orientation indicator. The above study warrants the use of  radio core to optical continuum luminosity
ratio as a reliable proxy for quasar orientation. 
 
 \citet{Shankar2008} found an inverse correlation between the BAL fraction and radio luminosity where BAL fraction is found  to be higher for low luminosity sources as compared to high luminosity sources.
 They invoked a simple orientation model to explain the  BAL fraction-radio luminosity trend. Their model assumes that some fraction of sources is boosted to higher luminosities due to relativistic beaming towards the observer. As BAL quasars are preferentially viewed close to the accretion disk plane, they are less beamed and less luminous on average. Thus, the trend of BAL fraction with radio luminosity is very much linked to the orientation of the jet with respect to the line of sight, and \citet{Shankar2008} essentially used this phenomenon to define radio luminosity as a proxy for quasar orientation.  The only caveat in their approach is that the fraction of radio sources with relativistic beaming is not well constrained. In our analysis, we alleviate this problem by adopting radio core dominance as the quasar orientation indicator.  By normalizing the radio core luminosity with the intrinsic power of the central engine, the amount of relativistic beaming is well constrained for sources in our sample. The most important result from our study is that the BAL fraction increases considerably for higher orientation angles of radio jets with respect to the line of sight. This trend is found in all the samples, irrespective of their morphological classes.  At low orientation angles, the BAL fraction is $\sim$ 20 percent which increases to $\sim$ 40 percent at high orientation angles.  In \citet{Shankar2008} study,  the BAL fraction for classical BALs reduced from 20 percent at faint luminosities to 8 percent at high luminosities. Adopting a broader definition of BAL quasars as given in \citet{Trump2006}, they found the fraction of radio BAL quasars to be 44 percent at low luminosities, reducing to 23 percent at high luminosities.  Therefore, our results are consistent with the findings of \citet{Shankar2008}. Interestingly, a Spearman-rank correlation analysis between the radio luminosity and the orientation parameter R$_I$ for sources in our sample results in a statistically significant correlation with a correlation coefficient of 0.7.

{   \cite{Gregg2006} too found a similar anti-correlation between the radio-loudness and strength of the BAL features. Fig.~3 in \cite{Gregg2006} shows the anti-correlation between balnicity index and radio-loudness for BAL quasars. The anti-correlation is stronger for the FR II–LoBAL quasars.  As the radio loudness parameter is similar to the orientation indicator R$_I$,  the sources with higher radio-loudness could indicate low-inclination angles to the jet axis due to core-boosting. Hence, the anti-correlation between balnicity index and radio-loudness is consistent with our results which favors a prominent role of orientation. However, \cite{Gregg2006} argued for the evolutionary model of BAL quasars due to the paucity of FR-II BAL quasars in a large sample of quasars. They argued that the emergence of radio jets to create FR II sources is frustrated by the obscuring BAL shroud in the quasar infancy.  In our study, we note that the fraction of BAL quasars in lobe-dominated sources is similar to core-dominated and resolved core quasars (See, Table.~\ref{tab:balf}). Among the BAL quasar sources with extended lobe-dominated morphology, our sample contains 49  doubles and 58  triples. Hence, we do not notice any paucity of FR-II BAL quasars in our sample. The lack of FR-II BAL quasars in earlier studies may arise due to adopting  (i) a  stringent definition of BAL quasars based on the balnicity index and (ii) radio loudness as a criteria for sample selection.}

Several studies have tried to probe the orientation of BAL quasars using radio spectral index \citep{Becker2000,Montenegro-Montes2008,Fine2011,DiPompeo2011a}. These studies have shown that BAL quasars have a wide range of radio spectral indices suggesting a full range of orientations. { \cite{Becker2000} identified 29 BAL quasars in the FIRST Bright Quasar Survey where two-thirds of sources had steep radio spectra, and the remaining third had flat spectra.} Earlier studies did not find any significant difference in the distribution of radio spectral indices for BAL and non-BAL quasars, probably due to small sample sizes.   Using a sample of 74 radio-selected BAL quasars, \citet{DiPompeo2011a} showed that  BAL quasars has an over-abundance of steep spectrum sources compared to non-BAL quasars, though both samples spanned a wide range of the spectral index. They concluded that BAL quasars are more often observed at high orientation angles compared to typical quasars. The abundance of steep spectrum sources among BAL quasars is again consistent with our finding of increased BAL fraction at high orientation angles. Additionally, the fact that we find an appreciable BAL fraction $\sim$ 20 percent, even at low orientation angles, is congruent with the wide range of radio spectral indices found in BAL quasars.

We argue that our analysis is superior to the previous studies of BAL quasar orientations for the following reasons: (a) As we work with the latest release of SDSS data, our BAL quasar sample sizes are considerably larger than the previous studies by at least 10 times, (b) As our analysis mainly uses  radio images which are readily available from large scale surveys like FIRST, our radio quasar sample sizes are also significantly larger than the samples used in radio spectral index studies, (c) The automatic classification of radio sources into different morphological classes has enabled us to study the variation of BAL fraction with orientation in these sub-classes, (d) By adopting the radio core to optical luminosity ratio, we have used the best available proxy for quasar orientation, and (e) We also show that the BAL fraction-orientation angle trend holds for different   BAL quasar sub-classifications and find a succession in the rate of BAL fraction increase between HiBALs, LoBALs and FeLoBALs.

It is easy to explain our results using a pure geometric model where BAL quasars are preferentially seen close to the accretion disk plane. The progression in the rate of BAL fraction increase between   HiBALs, LoBALs, and FeLoBALs probably points to a statistical progression in the distribution of orientation angles between these sources. However, there are several pieces of evidence that tell that the scenario is not that simple. In this study, we find an appreciable fraction of BAL quasars even at low orientation angles irrespective of the radio morphology and BAL sub-classification. The discovery of polar BAL quasars \citep{Zhou2006}, compactness of BAL quasars at high-resolution radio images \citep{Kunert2010}, presence of BAL quasars with convex radio spectral shapes like Giga-Hertz peaked sources \citep{Montenegro-Montes2008} are some of the results which challenge the pure orientation model. Even in our analysis, we see a tentative increase in the BAL fraction towards very low orientation angles pointing to the existence of polar BAL quasars. These results suggest that orientation plays a role, apparently a significant one, in the presence of BAL features but is inadequate to explain the whole class of BAL quasars.  A combination of orientation and evolution is probably the best way to explain the observed fraction BAL quasars.

\section{Conclusion}
\label{sec:conclusion}
The following are the point-wise conclusions of our analysis that probed the orientation dependence of BAL quasars using a sample of sources commonly available in SDSS and FIRST surveys.

\begin{enumerate}
    \item We built three CNN models  for classifying radio quasars  based on the morphology of their radio emission. The first model (broad classification) is used to classify the sources into Core, Jets, Doubles, and Triples. The other two models (sub-classification of Doubles and Triples) are used separately to sub-classify Doubles and Triples into core-dominated and lobe-dominated sources. All the three CNN models (Section \ref{sec:cnn}) show an F1 score >98$\%$.  \\
    
    \item  The average BAL fraction appears to be nearly identical for the morphological classes of resolved core, core-dominated, and cobe-dominated quasars.  This average BAL fraction $\sim$ 23 \% is consistent with BAL fractions inferred from the infrared and radio wavebands.\\
    
    \item We used orientation indicators such as the core-to-lobe flux density ratio (R) and core radio loudness parameter (R$_I$) to study the distribution of BALs at different inclination angles with respect to the symmetry axis of the accretion disk. It is seen that for all morphologies defined in this paper, the BAL fraction is high ($\sim$ 40 \%) at higher orientation angles and low ($\sim$ 20 \%) at low orientation angles.  This increase in BAL quasar fraction is independent of any bias from optical or radio selection effects.\\
    
    \item We used the sources from \cite{Trump2006} catalog to study the distribution of orientation indicator log(R$_I$) amongst HiBALs, LoBALs, and FeLoBALs. The trend for BAL fraction to increase at high orientation angles is seen in \cite{Trump2006} catalog sources. Additionally, the rate of increase of BAL fraction is maximum for FeLoBALs, intermediate for LoBALs, and minimum for HiBALs. \\
    
    \item The increase of BAL fraction at high orientation angles statistically favors the geometric model of BAL quasars. However, the presence of significant fraction of BAL quasars at all orientation angles, tentative evidence of the presence of polar BAL quasars in our analysis may point to a scenario where a pure geometric model is insufficient to explain the nature of BAL quasars completely.  
    
\end{enumerate}
The present study  adds to the growing evidence of a more prominent role of orientation in explaining the BAL phenomena, but at the same time,   also hints at a significant role of evolution. { We also have started a detailed study of lobe-dominated BAL quasars in our sample.}
Upcoming radio sky surveys such as the LOFAR survey, VLASS, and  SKA will increase the number of radio sources by several times.  This will significantly improve the statistics for extreme R$_I$ bins. With these datasets, it will be desirable to unambiguously establish the existence of  polar BAL quasars at low orientation angles.

\section*{Acknowledgements}
{ We thank the anonymous referee  for the feedback which has significantly helped to improve the paper.}
MV acknowledges support from DST-SERB in the form of core research grant(CRG/2020/1657).

Funding for the Sloan Digital Sky 
Survey IV has been provided by the 
Alfred P. Sloan Foundation, the U.S. 
Department of Energy Office of 
Science, and the Participating 
Institutions. 

SDSS-IV acknowledges support and 
resources from the Center for High 
Performance Computing  at the 
University of Utah. The SDSS 
website is www.sdss.org.

SDSS-IV is managed by the 
Astrophysical Research Consortium 
for the Participating Institutions 
of the SDSS Collaboration including 
the Brazilian Participation Group, 
the Carnegie Institution for Science, 
Carnegie Mellon University, Center for 
Astrophysics | Harvard \& 
Smithsonian, the Chilean Participation 
Group, the French Participation Group, 
Instituto de Astrof\'isica de 
Canarias, The Johns Hopkins 
University, Kavli Institute for the 
Physics and Mathematics of the 
Universe (IPMU) / University of 
Tokyo, the Korean Participation Group, 
Lawrence Berkeley National Laboratory, 
Leibniz Institut f\"ur Astrophysik 
Potsdam (AIP),  Max-Planck-Institut 
f\"ur Astronomie (MPIA Heidelberg), 
Max-Planck-Institut f\"ur 
Astrophysik (MPA Garching), 
Max-Planck-Institut f\"ur 
Extraterrestrische Physik (MPE), 
National Astronomical Observatories of 
China, New Mexico State University, 
New York University, University of 
Notre Dame, Observat\'ario 
Nacional / MCTI, The Ohio State 
University, Pennsylvania State 
University, Shanghai 
Astronomical Observatory, United 
Kingdom Participation Group, 
Universidad Nacional Aut\'onoma 
de M\'exico, University of Arizona, 
University of Colorado Boulder, 
University of Oxford, University of 
Portsmouth, University of Utah, 
University of Virginia, University 
of Washington, University of 
Wisconsin, Vanderbilt University, 
and Yale University.

\section*{Data Availability}

The CNN models and catalog of the 16491 sources with coordinates and radio loudness parameters can be found at https://github.com/nairakhils/QuasarMorphCNN. DR16Q and \citet{Trump2006} catalog information are available from their respective papers. FIRST images can be downloaded using the FIRST cutout server link https://third.ucllnl.org/cgi-bin/firstcutout.



\bibliographystyle{mnras}
\bibliography{example} 



\appendix

\section{Examples of sources classified using CNN models}

\begin{figure*}
	\includegraphics[scale=0.16]{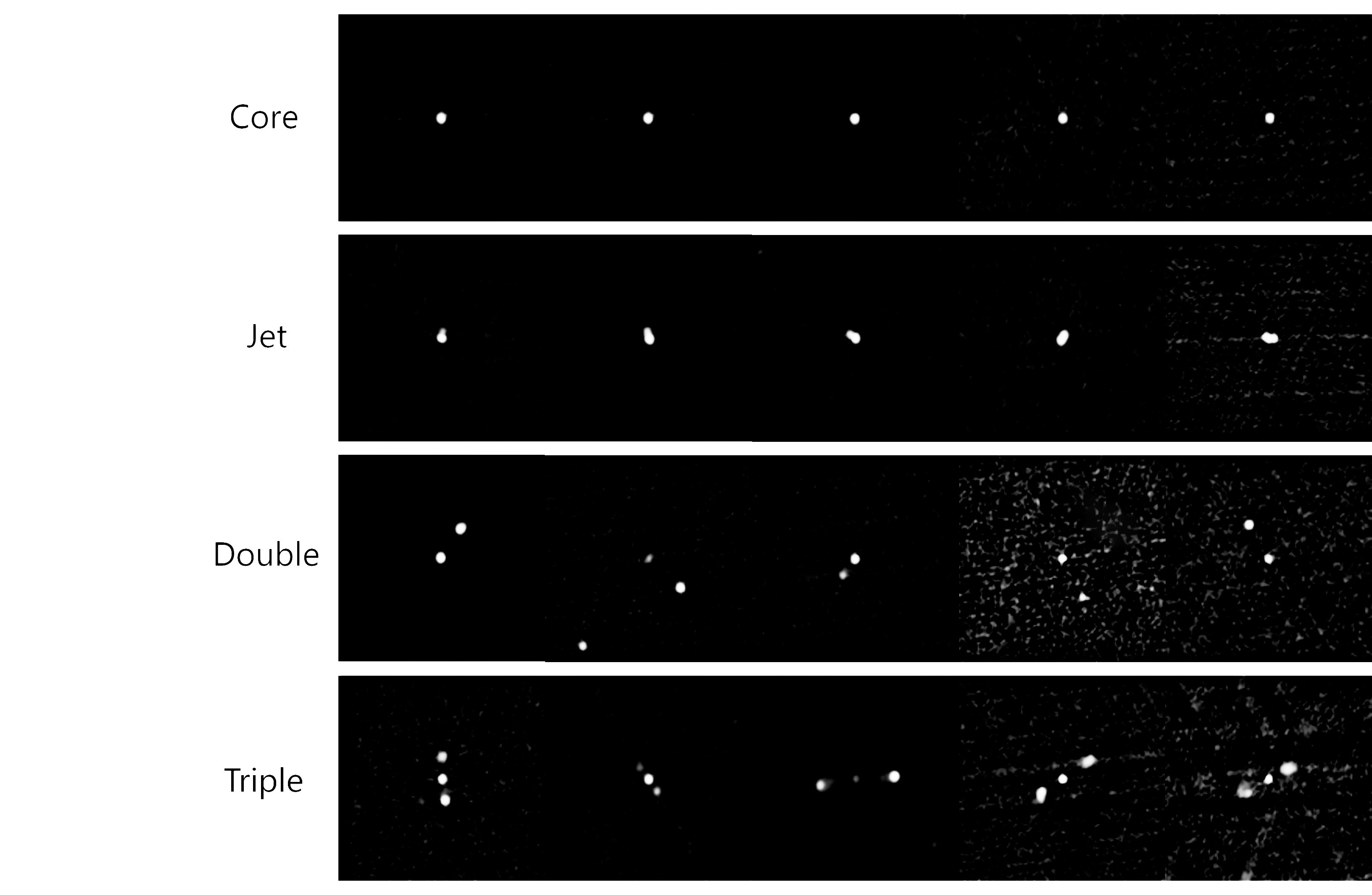}
    \caption{Examples of morphologically classified broad classes of radio AGN using CNN model defined in section \ref{sec:cnn}.}
    \label{fig:exmain}
\end{figure*}

In Fig.~\ref{fig:exmain}, we can see five examples each of all the morphological classes predicted by the CNN model defined in section \ref{sec:cnn}. 

\begin{figure*}
	\includegraphics[scale=0.12]{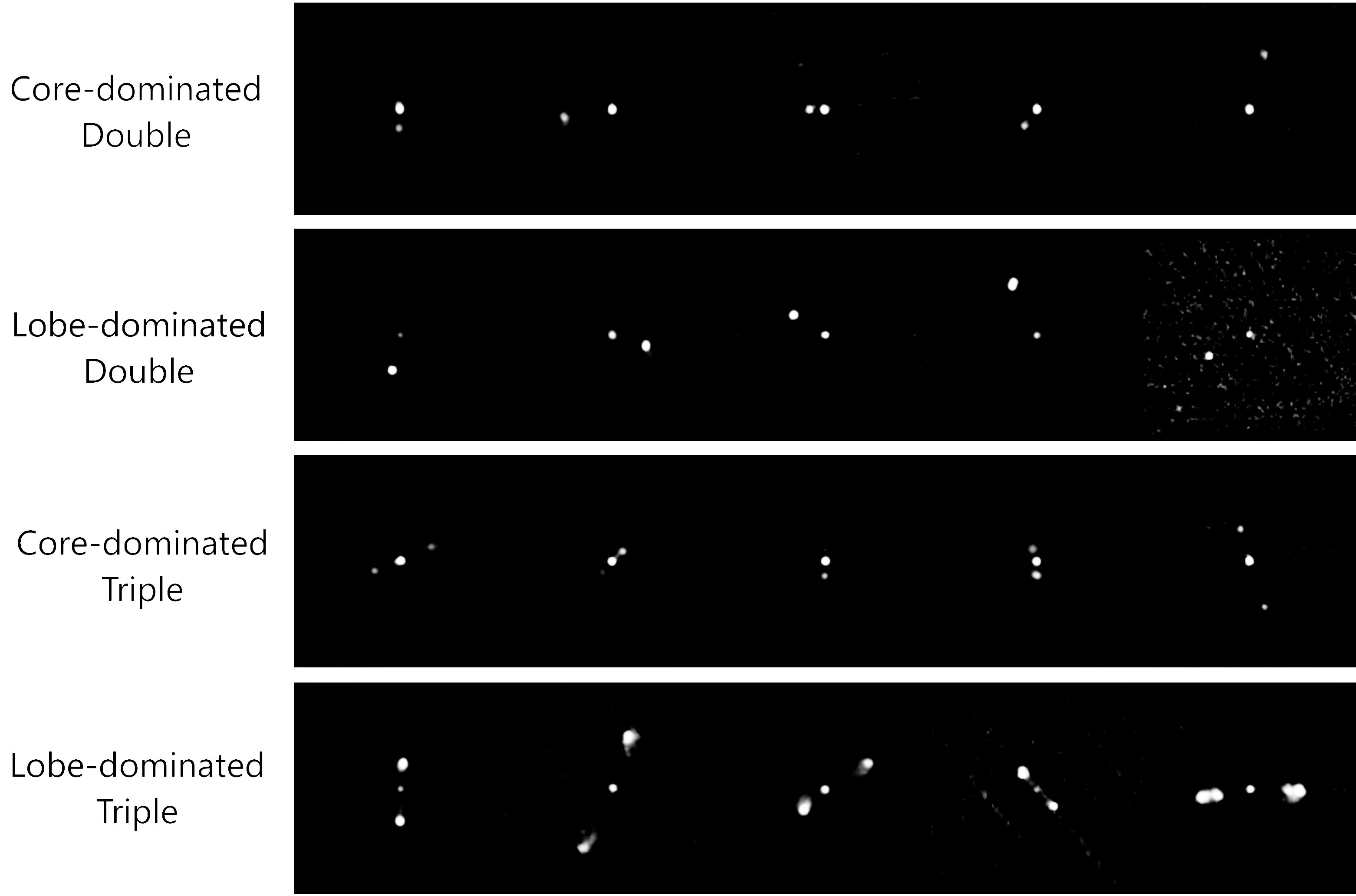}
    \caption{Examples of morphologically classified sub-classes of radio AGN using CNN models defined in section \ref{sec:cnn}.}
    \label{fig:exsub}
\end{figure*}

\begin{figure*}
	\includegraphics[scale=0.75]{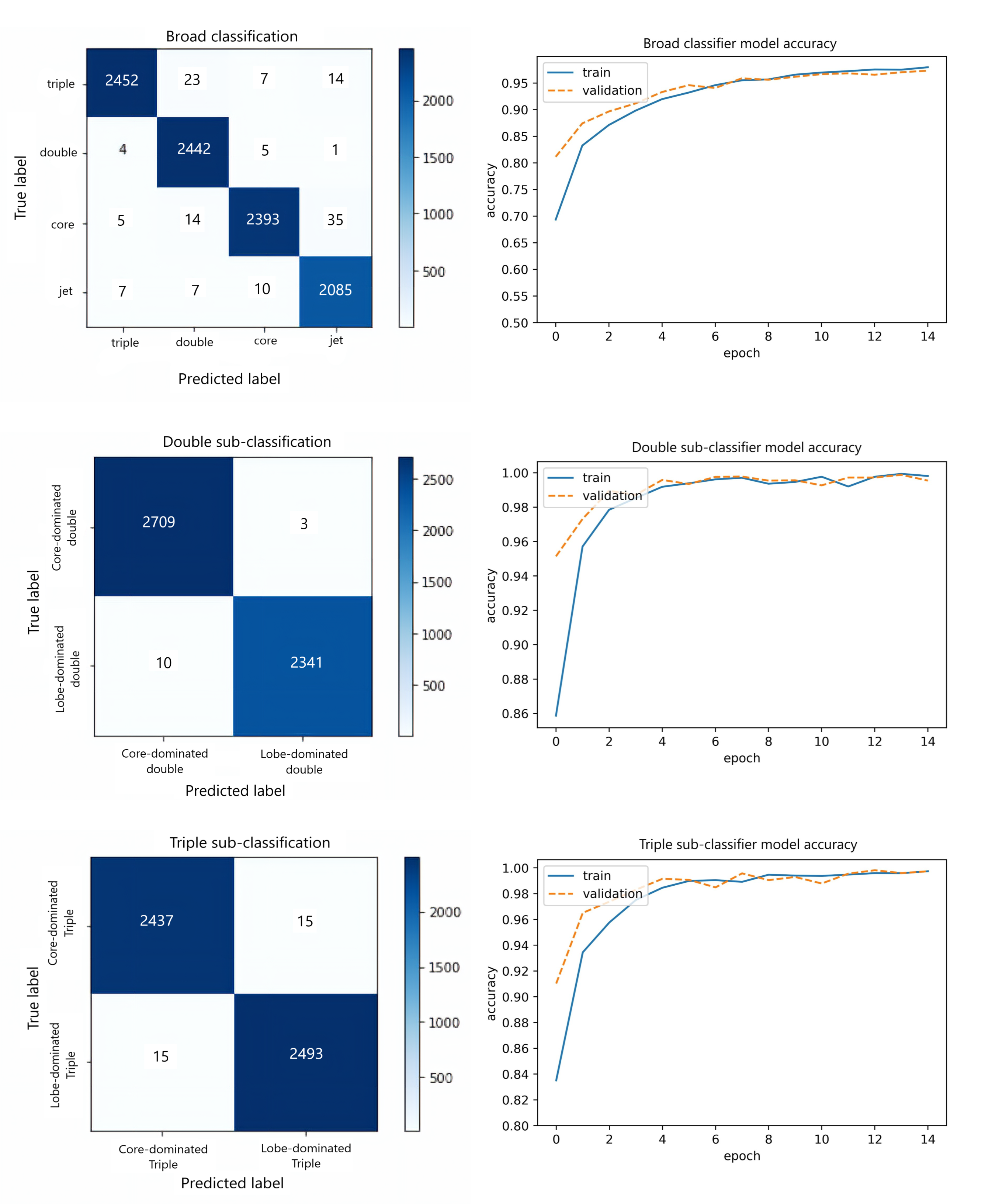}
    \caption{The confusion matrix and learning curves for the three CNN models (section~\ref{sec:cnn}): a) Broad classifier model (top panel), b) Double sub-classifier model (middle panel) and c) Triple sub-classifier model (bottom panel) are shown in this figure. The confusion matrices show the number of sources for each true/predicted label and the learning curves show the model's learning accuracy for the training and validation set.}
    \label{fig:cnnacc}
\end{figure*}

 We keep a track of the training and validation accuracy of the models at each epoch. This allows us to visualize the learning curve of our model. Right panels of Fig~\ref{fig:cnnacc} shows the learning curve for the broad classifier (top), double's sub-classifier (middle) and triple's sub-classifier (bottom), respectively. Left panels of Fig~\ref{fig:cnnacc} shows the corresponding  confusion matrices.

\FloatBarrier

\bsp	
\label{lastpage}
\end{document}